\documentclass[12pt]{article}
\usepackage{amssymb,esvect,amsmath,graphicx,latexsym,amsthm,slashed,eso-pic,hyperref,rotating,epsfig,verbatim,mathrsfs}
\usepackage{xcolor}
\usepackage{slashed}
\usepackage{multirow}
\usepackage[utf8]{inputenc}
\graphicspath{{./}{./Figs/}{./Figs/old/}}

\def\beq{\begin{eqnarray}}
\def\eeq{\end{eqnarray}}
\def\bea{\begin{eqnarray}}
\def\eea{\end{eqnarray}}

\def\gev{\, {\rm GeV}}
\def\mev{\, {\rm MeV}}

\newcommand{\gsim}{\lower.7ex\hbox{$\;\stackrel{\textstyle>}{\sim}\;$}}
\newcommand{\lsim}{\lower.7ex\hbox{$\;\stackrel{\textstyle<}{\sim}\;$}}

\newcommand{\nnmb}{\nonumber}

\newcommand{\lrf}[2]{\left(\frac{#1}{#2}\right)}

\newcommand{\lag}{\mathscr{L}}

\newcommand{\kev}{\,\mathrm{keV}}
\newcommand{\ev}{\, {\rm eV}}

\newcommand{\ngam}{\mathcal{N}_{\gamma}}

\newcommand{\eg}{E_{\gamma}}

\newcommand{\naa}{\mathcal{N}_{a}}
\newcommand{\nbb}{\mathcal{N}_{b}}

\newcommand{\bfgam}{\bar{f}_{\gamma}}

\newcommand{\lcdm}{\Lambda\text{CDM}}
\newcommand{\vbl}{B\!-\!L}
\newcommand{\vet}{L_e\!-\!L_{\tau}}
\newcommand{\vem}{L_e\!-\!L_\mu}
\newcommand{\vmt}{L_{\mu}\!-\!L_{\tau}}

\begin{document}

\begin{titlepage}
\noindent
\begin{center}
  \begin{Large}
    \begin{bf}
Cosmological Bounds on sub-GeV Dark Vector Bosons\vspace{0.2cm}\\ 
from Electromagnetic Energy Injection
     \end{bf}
  \end{Large}
\end{center}
\vspace{0.2cm}
\begin{center}
\begin{large}
John Coffey$^{(a)}$, 
Lindsay Forestell$^{(b,c)}$,\\ 
David E. Morrissey$^{(c)}$, and Graham White$^{(c)}$
\end{large}
\vspace{1cm}\\
\begin{it}
(a) Department of Physics and Astronomy, University of Victoria,\\
Victoria, BC V8P 5C2, Canada
\vspace{0.3cm}\\
(b) Department of Physics and Astronomy, University of British Columbia,\\ 
Vancouver, BC V6T 1Z1, Canada\vspace{0.3cm}\\
(c) TRIUMF, 4004 Wesbrook Mall, Vancouver, BC V6T 2A3, Canada
\vspace{0.5cm}\\
email: 
\emph{\texttt{jwcoffey@uvic.ca}},
\emph{\texttt{lmforest@phas.ubc.ca}},\\
\emph{\texttt{dmorri@triumf.ca}}, 
\emph{\texttt{gwhite@triumf.ca}}
\vspace{0.2cm}
\end{it}
\end{center}
\center{\today}

\begin{abstract}
New dark vector bosons that couple very feebly to regular matter can be
created in the early universe and decay after the onset of
big bang nucleosynthesis~(BBN) or the formation of the cosmic microwave
background~(CMB) at recombination.  The energy injected
by such decays can alter the light element abundances or modify the
power and frequency spectra of the CMB.
In this work we study the constraints implied by these effects
on a range of sub-GeV dark vectors including the kinetically mixed
dark photon, and the $\vbl$, $\vem$, $\vet$, and $\vmt$ dark $U(1)$ bosons.
We focus on the effects of electromagnetic energy injection, and 
we update previous investigations of dark photon and other dark vector
decays by taking into account non-universality in the photon cascade 
spectrum relevant for BBN and the energy dependence of the ionization 
efficiency after recombination in our treatment of modifications to the CMB.

\end{abstract}

\end{titlepage}

\setcounter{page}{2}


\section{Introduction\label{sec:intro}}

  Direct measurements of the cosmos give strong evidence for 
a standard $\lcdm$ cosmology containing the elementary
particles and forces predicted by 
the Standard Model~(SM)~\cite{Kolb:1990vq,Dodelson:2003ft}.
Observations of the cosmic microwave background~(CMB)~\cite{Hinshaw:2012aka,Aghanim:2018eyx},
baryon acoustic oscillations~(BAO)~\cite{Eisenstein:2005su,Cole:2005sx,Alam:2016hwk}, and supernovae distances~\cite{Riess:1998cb,Perlmutter:1998np,Scolnic:2017caz}
fix the parameters of the $\lcdm$ model very precisely.
The success of big bang nucleosyntheis~(BBN) at predicting the light element
abundances (notwithstanding the lithium puzzles) provides further
support for a $\lcdm$ cosmology with SM particles and interactions going back
even earlier in time, up to temperatures on the order of a few MeV~\cite{Schramm:1977ne,Walker:1991ap,Sarkar:1995dd,Iocco:2008va,Jedamzik:2009uy,Pospelov:2010hj,Cyburt:2015mya,Fields:2019pfx}.

  Tests of the SM+$\lcdm$ paradigm also put stringent constraints
on a broad range of new physics beyond the SM and deviations from the 
standard cosmology.  In particular, processes that inject energy
into the cosmological plasma can modify BBN by altering the ratio of 
neutrons to protons or destroying/creating light elements.
These considerations have been used to constrain new (effective) light degrees 
of freedom~\cite{Sarkar:1984tt,Cyburt:2004yc,Ho:2012ug,Boehm:2013jpa,Nollett:2013pwa,Ishida:2014wqa,Kamada:2015era,Kamada:2018zxi,Escudero:2018mvt,Escudero:2019gzq,Berlin:2019pbq,Sabti:2019mhn}, 
the minimum temperature of reheating or periods of non-standard cosmological 
evolution~\cite{Kawasaki:1999na,Kawasaki:2000en,Hannestad:2004px},
and the decays~\cite{Ellis:1984er,Juszkiewicz:1985gg,Dimopoulos:1987fz,Reno:1987qw,Dimopoulos:1988ue,Ellis:1990nb,Moroi:1993mb,Kawasaki:1994af,Cyburt:2002uv,Jedamzik:2004er,Kawasaki:2004qu,Jedamzik:2006xz,Kawasaki:2008qe,Kawasaki:2017bqm} 
or annihilations~\cite{Frieman:1989fx,Hisano:2008ti,Hisano:2009rc,Kawasaki:2015yya,Depta:2019lbe} of particles in the early universe.
Related bounds can be derived on late-time energy injection from the
deviations induced in the CMB frequency~\cite{Hu:1992dc,Hu:1993gc}
and power spectra~\cite{Adams:1998nr,Chen:2003gz,Padmanabhan:2005es,Zhang:2007zzh,Slatyer:2009yq,Huetsi:2009ex,Hutsi:2011vx,Dutra:2018gmv}.

  In this work we apply and extend these methods to derive cosmological 
bounds from BBN and the CMB on a range of sub-GeV dark vector bosons.  
New vector bosons are ubiquitous in extensions of the SM, including
grand unification~\cite{Slansky:1981yr,Hewett:1988xc,Leike:1998wr}, 
superstring theory~\cite{Blumenhagen:2005mu,Douglas:2006es,Blumenhagen:2006ci}, 
and theories of dark matter~\cite{Fayet:2007ua,Pospelov:2007mp,ArkaniHamed:2008qn}.
By definition, dark vector bosons couple only very feebly to the SM,
and can thus be much lighter than the electroweak scale without having
been discovered yet~\cite{Pospelov:2008zw,Bjorken:2009mm,Alexander:2016aln}.
For very small couplings, dark vectors are very difficult to produce
in the laboratory and can develop macroscopically long lifetimes.
Thermal cosmological production is also reduced in this regime and the
dark vectors might never develop a full thermal abundance.
Even so, dark vector bosons can still be created 
in the early universe through 
direct reheating~\cite{Berezhiani:1995am,Adshead:2016xxj,Hardy:2017wkr,Adshead:2019uwj}, 
freeze-in processes~\cite{Hall:2009bx,Cheung:2010gj,Chu:2011be,Chu:2013jja,Bernal:2017kxu}, 
or the decays or annihilations of heavier dark states~\cite{Hooper:2012cw,Hasenkamp:2012ii}.
The later decays of the vector bosons then inject energy into the 
cosmological medium, possibly modifying the light element abundances
and the CMB relative to SM+$\lcdm$~\cite{Fradette:2014sza,Berger:2016vxi}. 

We investigate a range of light (but massive) U(1) vector bosons
including a dark photon~(DP) that connects to the SM exclusively through 
gauge kinetic mixing with hypercharge~\cite{Okun:1982xi,Holdom:1985ag},
as well as the anomaly-free combinations 
$\vbl$~\cite{Heeck:2014zfa,Jeong:2015bbi}, 
$\vem$~\cite{Wise:2018rnb}, 
$\vet$~\cite{Wise:2018rnb}, 
and $\vmt$~\cite{Altmannshofer:2014pba,Kaneta:2016uyt} 
with only very tiny direct couplings to the SM.
Our focus is on the mass range $m_V \in [1\,\mev,\,1\,\gev]$ over
which these dark vectors have significant decay fractions to
electromagnetic channels, and we concentrate on the impact of 
the electromagnetic component of the energy injected by the decays
which is typically the most important effect for lifetimes greater than
$\tau_V\gtrsim 10^4\,\mathrm{s}$.

  This study expands on previous investigations of cosmological bounds on
dark photons in two ways~\cite{Fradette:2014sza,Berger:2016vxi}.  
First, applying the technology developed in Ref.~\cite{Forestell:2018txr}, 
we compute the full electromagnetic cascade induced by dark vector decays, 
including some new effects relative to 
Refs.~\cite{Fradette:2014sza,Berger:2016vxi}.  From this we derive
the resulting photon cascade spectrum that is essential for determining 
the impact of electromagnetic energy injection on the light element abundances.  
Our calculation also differs from the more common approach of using the 
universal photon spectrum defined in Ref.\cite{Cyburt:2002uv}, 
based on the results of Refs.~\cite{Ellis:1990nb,Agaronyan:1983xx,Svensson:1990pfo,Protheroe:1994dt,Kawasaki:1994sc}, 
to derive the cascade photon spectrum.
The universal spectrum method provides a very good description of 
the photon cascade produced by the interactions of highly-energetic 
electromagnetic primaries ($E \gtrsim 100\,\gev$) with the cosmological 
background prior to recombination, and has the major simplifying feature 
that it only depends on the total amount of electromagnetic energy injected 
rather than the detailed primary injection spectrum.
However, it was shown in Refs.~\cite{Forestell:2018txr,Poulin:2015woa,Poulin:2015opa,Hufnagel:2018bjp} 
that lower-energy electromagnetic injection near or below 
the GeV scale can lead to cascade photon spectra that deviate significantly 
from the predictions of the universal spectrum, and that can depend on the type 
and spectrum of the energy~\cite{Forestell:2018txr}.

In light of this first point, the second way that we expand on previous work
is by computing the detailed primary electromagnetic injection spectra 
from sub-GeV vector boson decays~\cite{Coogan:2019qpu,Plehn:2019jeo}.  
We study the energy spectra of electrons and photons created by the decay modes 
$V\to \{e^+e^-,\,\mu^+\mu^-,\,\pi^+\pi^-,\,\pi^+\pi^-\pi^0,\,\pi^0\gamma\}$, 
which are expected to be the most important ones for the dark vectors of 
interest~\cite{Ilten:2018crw,Bauer:2018onh}. 
We then apply these injection spectra to computing the full electromagnetic 
cascade spectrum produced by a given dark vector decay for our BBN studies
or use them to calculate limits from the CMB.

The outline of this paper is as follows. In Sec.~\ref{sec:vdec} 
we compute the decay rates and fractions of the dark vector species
and we find the energy spectra of the resulting photons and electrons.
Next, in Sec.~\ref{sec:bbn} we study the effects
of electromagnetic energy injected by dark vector decays on the light elements
and use current data to set limits on the pre-decay dark vector densities 
as a function of their lifetime and mass.  We investigate the related effects
of dark vector decays on the power and frequency spectra of the CMB
in Sec.~\ref{sec:cmb}.  In Sec.~\ref{sec:cosmo} we review the calculation
of the density of dark vectors from thermal production in the early universe,
and we find the constraints from BBN and the CMB on the masses and couplings 
of dark vectors for the corresponding thermal yields.  
Finally, Sec.~\ref{sec:conc} is reserved for our conclusions.
Some additional details on the departure from electromagnetic universality
in BBN and when it occurs are presented in Appendix~\ref{sec:appb}.

\section{Dark Vector Bosons and their Decays\label{sec:vdec}}

  We begin by reviewing the dominant decay channels for the sub-GeV
dark vector bosons of interest: kinetically mixed dark photon~(DP), 
$\vbl$, $L_e\!-\!L_\mu$, $L_e\!-\!L_\tau$, $L_\mu\!-\!L_\tau$.
In all cases we assume the dark vectors decay exclusively to the SM, 
with no additional exotic decay channels available.
With the decay fractions in hand, we turn next to computing
the electron and photon spectra they produce in the rest frame
of the dark vector relevant for computing cosmological bounds on them.

\subsection{Vector Boson Decay Branching Fractions}

  The sub-GeV dark vectors under consideration interact with the SM
primarily through gauge kinetic mixing for the dark photon~(DP),
\beq
\lag \ \supset \ -\frac{\epsilon}{2}\,F_{\mu\nu}V^{\mu\nu} 
\label{eq:dpkin}
\eeq
with kinetic mixing parameter $\epsilon$,
and via direct gauge coupling to the SM for the rest,
\beq
\lag \ \supset \ - g_V\bar{f}\gamma^{\mu}(Q_{L}P_L+Q_{R}P_R)f\,V_{\mu}
\eeq
where $g_V\ll 1$ and we normalize to $|Q_{L}| = 1$ for leptons.  
The direct-coupling dark vectors will also have a gauge kinetic mixing
with the photon but we assume (self-consistently) that it is sub-leading
relative to the direct gauge interaction.
We assume further that the dark vectors obtain a mass $m_V$ in the MeV--GeV range
from the Higgs~\cite{Schabinger:2005ei} 
or Stueckelberg~\cite{Stueckelberg:1900zz,Feldman:2007wj} mechanisms,
and that no decay channels to light SM-singlet states are available.
In particular, for the direct-coupling vectors we assume as 
in Ref.~\cite{Bauer:2018onh} that the light neutrinos are Majorana 
and mostly left-handed 
(which requires a spontaneous breaking of the underlying gauge invariance), 
and that any SM singlets are heavy enough to be neglected.

  With these couplings, the perturbative decay widths to SM fermions $f$ are
\beq
\Gamma(V\to f\bar{f}) = 
N_c\frac{\alpha_{V}}{6}\,m_V\,
\sqrt{1-\lrf{2m_f}{m_V}^2}\;
\left(Q_L^2+Q_R^2 + \left[6Q_LQ_R-(Q_L^2+Q_R^2)\right]\frac{m_f^2}{m_V^2}\right) 
\eeq
where $N_c$ is the number of QCD colours of the fermion, 
$\alpha_{V} = \epsilon^2\alpha$ for the dark photon (with $Q_{L,R} = Q_{em}$),
and $\alpha_{V} = g_V^2/4\pi$ for the direct vectors (with $Q_R=0$
for the light neutrinos given the assumptions stated above).

  For sub-GeV vectors, we must also consider decays of the vector to 
specific hadronic states.
Since these decays are strongly suppressed for the leptonic vectors, we only
include them for the DP and $\vbl$ vectors here.
Our approach to describing hadronic dark vector decays uses 
a combination of the data-driven methods 
of Refs.~\cite{Fradette:2014sza,Ilten:2018crw,Bauer:2018onh} 
and the analytic description based on vector-meson-dominance~(VMD) 
of Ref.~\cite{Tulin:2014tya}.
In the VMD picture~\cite{Sakurai:1960ju,Fujiwara:1984mp,Bando:1984ej,OConnell:1995nse}, 
which also guides the methods of Refs.~\cite{Fradette:2014sza,Ilten:2018crw,Bauer:2018onh}, 
the $\rho$ and $\omega$ vector mesons
are treated as gauge bosons of a hidden $U(2)$ flavour symmetry.\footnote{Note
that since we focus on sub-GeV dark vectors, it is sufficient to consider
flavour $U(2)$ with only up and down quarks and $\rho$ and $\omega$ mesons
rather than the larger approximate flavour $U(3)$.}
Dark vector decays to hadrons in this picture occur through direct couplings
-- the DP to the electromagnetic current and the $\vbl$ vector
to the baryonic current -- as well as through an induced kinetic mixing
with the $\rho$ and $\omega$ with approximate 
strength~\cite{Tulin:2014tya,Ilten:2018crw}
\beq
\epsilon_{VMD} \ \simeq \ 2\,\text{tr}(t_{A}\,Q_V)\,\frac{g_V}{g}
\ = \ 
\frac{g_V}{g}\times\left\{
\begin{array}{ccl}
1&;&\text{DP -- }\rho\\
1/3&;&\text{DP -- }\omega\\
0&;&(\vbl)\text{ -- }\rho\\
2/3&;&(\vbl)\text{ -- }\omega
\end{array}\right.
\eeq
where $t_{A} = \text{diag}\big(1/2,\,\pm 1/2\big)$ for $A= \omega,\,\rho$
is the $U(2)$ generator,
$g \simeq \sqrt{3\times 4\pi}$,
$Q_V = \text{diag}\big(2/3,\,-1/3\big)$ for DP 
and $Q_V = \text{diag}\big(1/3,\,1/3\big)$ for $\vbl$,
and $g_V = -e\epsilon$ for the DP.

  The leading hadronic decay mode of the dark photon is usually
$V\to \pi^+\pi^-$. In the VMD picture, it arises from a combination
of the direct countribution of the charged pions to the electromagnetic
current and the induced kinetic mixing with the $\rho$.
The corresponding decay width is
\beq
\Gamma(\text{DP}\to\pi^+\pi^-) \ = \ \frac{\alpha_V}{12\pi}\left[
1-\lrf{2m_{\pi^+}}{m_V}^2\right]^{3/2}|F_{\pi\pi}(m_V)|^2 \ ,
\eeq
where the leading-order form factor is~\cite{OConnell:1995nse} 
\beq
F_{\pi\pi} \ \simeq \
1 - \frac{g_{\rho\pi\pi}}{g}\frac{s}{s-m_{\rho}^2+im_{\rho}\Gamma_{\rho}(s)} \ ,
\label{eq:fpipi}
\eeq
with $g_{\rho\pi\pi} \simeq g$, $m_{\rho} = 775.25\,\mev$, 
and $\Gamma_{\rho}(s=m_{\rho}) = 149\,\mev$.  The true form factor 
is modified relative to Eq.~\eqref{eq:fpipi} by the energy dependence 
of the $\rho$ self-energy as well as $\rho$-$\omega$ mixing.  
In our analysis we follow Ref.~\cite{Ilten:2018crw} and extract 
the form factor from $e^+e^-\to \pi^+\pi^-(\gamma)$ measurements by 
the BABAR experiment~\cite{Lees:2012cj}.
Note that $\rho$-$\omega$ mixing also allows the $\vbl$ vector to
decay to $\pi^+\pi^-$ but we do not include the effect because the branching
fraction for this channel is always less than a couple percent.

Mixing with the $\omega$ allows the DP and $\vbl$ vectors to
decay to $\pi^+\pi^-\pi^0$ and $\pi^0\gamma$, with the effect being 
strongest near the $\omega$ mass.  
The width for $V\to \pi^0\gamma$ is~\cite{Tulin:2014tya}
\beq
\Gamma(V\to \pi^0\gamma) \ = \
\Big[2\text{tr}(t_AQ_V)\Big]^2\,
\frac{3\,\alpha_V}{128\pi^3}\frac{m_V^3}{f_\pi^2}\,
|F_{\omega}(m_V^2)|^2 \ .
\eeq
For the decay $V\to \pi^+\pi^-\pi^0$, the width is~\cite{Tulin:2014tya}
\beq
\Gamma(V\to \pi^+\pi^-\pi^0) \ = \ 
\Big[2\text{tr}(t_AQ_V)\Big]^2\,
\frac{3\,\alpha_V}{16\pi^4}\lrf{g^2}{4\pi}^{\!2}
\frac{m_V}{f_\pi^2}\,\mathcal{I}(m_V^2)
|F_{\omega}(m_V^2)|^2
 \ ,
\label{eq:3pi}
\eeq
where $g^2/4\pi \simeq 3.0$, the relevant form factor is
\beq
F_{\omega}(s) = \frac{s}{s-m_{\omega}^2+im_{\omega}\Gamma_{\omega}}
\eeq
with $m_{\omega} = 782.65\,\mev$ and $\Gamma_{\omega} = 8.49\,\mev$,
and~\cite{Fujiwara:1984mp} 
\beq
\mathcal{I}(m_V^2)  &=&  \int_{m_{\pi}}^{E_*}\!\!dE_+\!\int_{E_1}^{E_2}\!\!dE_-\;
\left[\vec{p}_+^{\;2}\vec{p}_-^{\;2} - (\vec{p}_+\cdot\vec{p}_-)^2\right]
\bigg|\frac{1}{(p_++p_-)^2-m_{\rho}^2+im_{\rho}\Gamma_{\rho}}
\label{eq:3piint}\\
&&
\hspace{1.1cm}
+ \frac{1}{(p_-+p_0)^2-m_{\rho}^2+im_{\rho}\Gamma_{\rho}}
+\frac{1}{(p_0+p_+)^2-m_{\rho}^2+im_{\rho}\Gamma_{\rho}}
\bigg|^2 \ ,
\nnmb
\eeq
where $E_{\pm}$ are the energies of the outgoing charged pions 
in the decay frame, $p_{i} = (E_i,\vec{p}_i)$ are the pion 4-momenta 
whose components can all be expressed in terms of $E_{\pm}$ 
by momentum conservation, and the integration region covers
\beq
E_* = \frac{m_V^2-3m_{\pi}^2}{2m_V} \ ,~~~~~
E_{1,2} = \frac{1}{2}\left(
m_V - E_+ \mp \lVert\vec{p}_+\rVert
\sqrt{\frac{m_V^2-2m_VE_+-3m_{\pi}^2}{m_V^2-2m_VE_++m_{\pi}^2}}\;
\right) \ .
\eeq

\begin{figure}[ttt]
 \begin{center}
         \includegraphics[width = 0.47\textwidth]{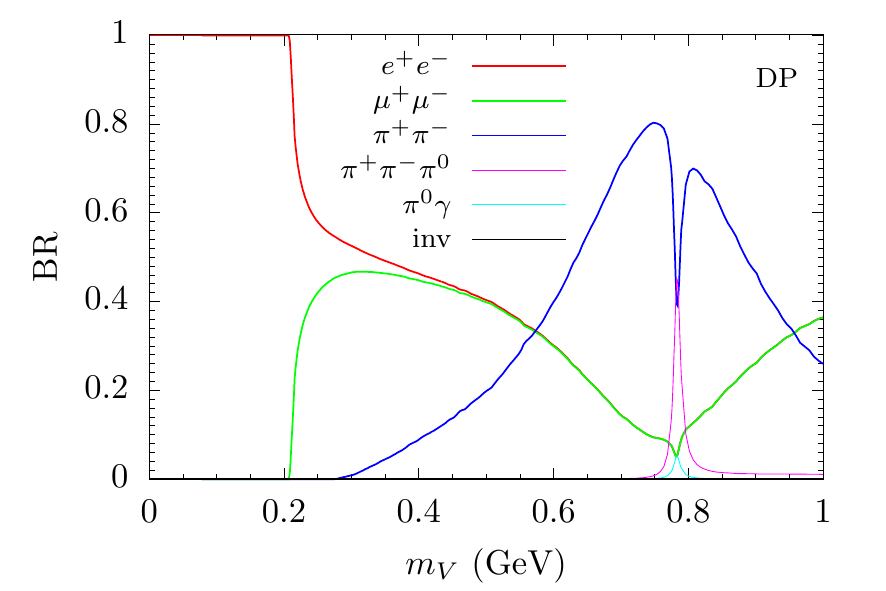}
         \includegraphics[width = 0.47\textwidth]{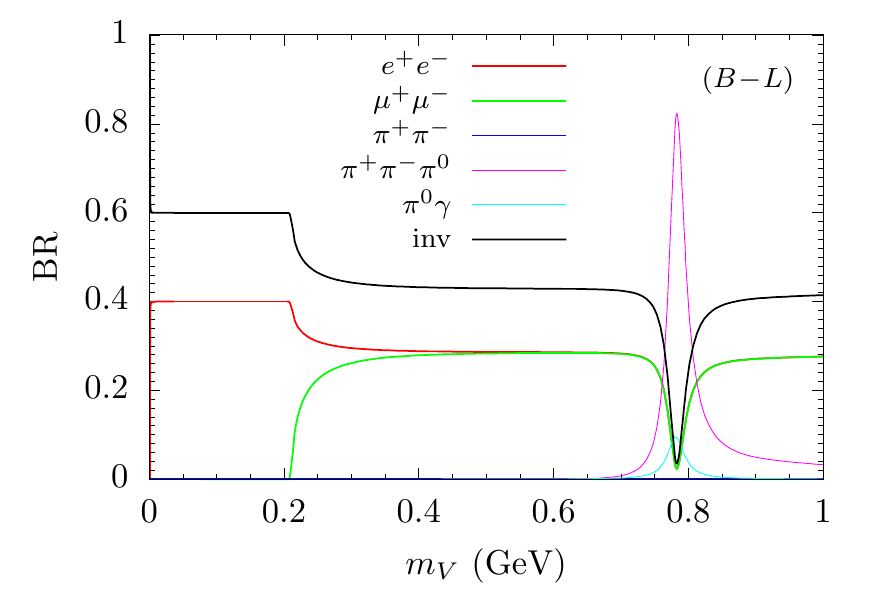}
 \end{center}
\vspace{-0.7cm}
 \caption{Decay ranching ratios~(BR) as a function of the vector mass $m_V$
for a sub-GeV dark photon~(left), or $\vbl$~(right) vector boson.}
 \label{fig:vprodbr}
 \end{figure}

In Fig.~\ref{fig:vprodbr} we show the branching fractions~(BR) 
of the dark photon~(left)
and $\vbl$ vector~(right) over the mass range of interest.
The hadronic resonance structures are evident near the $\rho$ and $\omega$
poles, while the leptonic modes dominate at lower masses.
The branching fractions of the leptonic vectors can also be obtained
straightforwardly and are mostly determined by the relative charges
and open decay channels.  For the $\vmt$ vector,
decays below the muon threshold are dominated by neutrinos
with only a $\text{BR} \lesssim 3\times 10^{-5}$ fraction to $e^+e^-$
(via induced kinetic mixing)~\cite{Bauer:2018onh}.
Hadronic decays of these vector bosons were also computed 
in Ref.~\cite{Bauer:2018onh}, and their branchings 
$\text{BR}_{had} \lesssim 10^{-3}$ are too small to have 
a relevant cosmological effect over the ranges we study.

\subsection{Electromagnetic Injection Spectra}

  The analysis above shows that the relevant decay channels of the dark vectors 
of interest are:
$e^+e^-$, $\mu^+\mu^-$, $\pi^+\pi^-$, $\pi^+\pi^-\pi^0$, $\pi^0\gamma$,
$\nu\bar{\nu}$.  These decays will all ultimately produce
a collection of photons, electrons, and neutrinos.  The resulting
energy spectra of photons and electrons from these decay modes, 
in the lab frame where the decaying vector boson is at rest,
are needed for our study of cosmological constraints to follow.
Together, the full spectra per decay are
\beq
\frac{dN_a}{dE} \ = \ \sum_{i}\text{BR}(V\to i)\,\frac{dN_a^{(i)}}{dE} \ ,~~~~~
a = \gamma,\,e,\,\nu
\eeq
where the sum runs over the decay modes $i= e^+e^-,\,\mu^+\mu^-,\;\ldots$,
and $e$ refers to the sum of electrons and positrons.\footnote{We treat
electrons and positrons as equivalent and we refer to them collectively
as electrons.} 
We compute the spectra for each of the exclusive decay modes here, 
concentrating on the electrons and photons produced at leading 
non-trivial order.  Our analysis is similar to that 
of Ref.~\cite{Coogan:2019qpu} with some differences and additions.
See also Ref.~\cite{Plehn:2019jeo} for an alternate approach
based on HERWIG~7~\cite{Bellm:2015jjp}.

\subsubsection*{$\mathbf{V\to e^+e^-}$}\vspace{-0.3cm}
This channel was considered previously in Ref.~\cite{Forestell:2018txr}.
The leading-order electron spectrum is trivial with two electrons,
each with energy $E = m_V/2$,
\beq
\frac{dN_e}{dE} = 2\;\delta(E-m_V/2) \ .
\eeq
In addition to electrons, 
photons can be produced from final-state-radiation~(FSR).
Following Ref.~\cite{Coogan:2019qpu}, we use the spectrum per decay
\beq
\left.\frac{dN_{\gamma}}{dE}\right|_{FSR} &=& \lrf{2}{m_V}\frac{\alpha}{\pi}\,
\frac{1}{(1+2\mu_f^2)\sqrt{1-4\mu_f^2}}\;\frac{1}{x}\;\Bigg(
\label{eq:fsr}
\\
&&
{}\hspace{0.5cm}
\left[1 + (1-x)^2 - 4\mu_f^2(x+2\mu_f^2)\right]
\ln\!\lrf{1+\mathcal{W}_f}{1-\mathcal{W}_f}\nnmb\\
&&
{}\hspace{2.3cm}
- \left[1 + (1-x)^2 -4\mu_f^2(1-x)\right]\,\mathcal{W}_f
\Bigg) \ ,
\nnmb
\eeq
with $x = E_{\gamma}/(m_V/2)$, $\mu_f = m_f/m_V$, 
and $\mathcal{W}_f = \sqrt{1-4\mu_f^2/(1-x)}$.
As shown in Ref.~\cite{Forestell:2018txr}, 
these FSR photons can have an important effect on BBN for sub-GeV injection.

\subsubsection*{$\mathbf{V\to \mu^+\mu^-}$}\vspace{-0.3cm}
At leading order, this channel ultimately produces two electrons 
and four neutrinos.  Beyond leading order, photons can be created
through FSR and radiative muon decays.

To obtain the leading-order energy spectrum of electrons in the 
vector boson lab frame, we focus on the dominant 
$\mu^-\to e^-\bar{\nu}_e\nu_{\mu}$ decay channel.
In the muon rest frame, for muon polarization $P \in [-1,1]$,
the electron energy spectrum is the well-known 
Michel form~\cite{Michel:1949qe,Tanabashi:2018oca}:
\beq
\left.\frac{dN_e}{dE'}\right|_{\mu} = \frac{2}{m_{\mu}}\,
2{y'}^2\left[(3-2y')+P(1-2y')\,\cos\vartheta\right]\,\Theta(1-y') \ ,
\label{eq:michel}
\eeq
where $y'=2E'/m_\mu$ and $\vartheta$ is the angle between the electron 
direction and the polarization axis in this frame.  
For anti-muon decays, the same spectrum applies but with the sign 
of $P$ reversed.  However, for $V\to\mu^+\mu^-$ the net effect of 
muon polarization cancels and this term can be neglected.
To get the lab frame spectrum we must boost the muon rest frame
with Lorentz factor $\gamma = m_V/2m_{\mu} = 1/\sqrt{1-\beta^2}$.  
Choosing the $z$ axis to lie along the muon boost direction
and assuming the electron is emitted at polar angle $\theta'$
from this axis in the muon rest frame,
the electron energy and momenta in the lab frame are
\beq
E = \gamma(E'+\beta\,p'\cos\theta') \ ,~~~
p_z = \gamma(p'\cos\theta' + \beta E') \ ,~~~
p_x = p'_x \ ,~~~
p_y = p'_y \ .
\eeq
Changing variables from $(E',\Omega')$ to $(E,\Omega')$ 
for solid angle $\Omega'$ in the muon frame, we obtain
the full lab frame distribution
\beq
\frac{dN_e}{dE} = 
\frac{2}{4\pi}\int\!d\Omega'\;
\frac{1}{\gamma(1+\beta\cos\theta'\,E'/p')}\:
\left.\frac{dN_e}{dE'}\right|_{\mu} \ ,
\label{eq:muboost}
\eeq
where $E' = E'(E,\Omega)$, the first term in the integrand is 
the relevant Jacobian factor, the factor of $1/4\pi$ normalizes
the solid angle integral, and the factor of two counts
the identical contributions from the $\mu^-$ and $\mu^+$ branches.

 The photon spectrum from muon decays is computed following 
Refs.~\cite{Coogan:2019qpu,Scaffidi:2016ind}
as a sum of FSR and radiative contributions,
\beq
\frac{dN_{\gamma}}{dE} = \left.\frac{dN_{\gamma}}{dE}\right|_{\text{FSR}} 
+ \left.\frac{dN_{\gamma}}{dE}\right|_{\text{rad}} \ .
\label{eq:mugam}
\eeq
This approach neglects interference between the two channels,
but this effect is expected to be very small due to the narrow width of the muon.
We use the expression of Eq.~\eqref{eq:fsr} with $m_f = m_{\mu}$
for the FSR part.  The radiative spectrum is the sum of contributions
from the $\mu^+$ and $\mu^-$ decays.  In the muon rest frame, both are equal
to~\cite{Coogan:2019qpu,Scaffidi:2016ind,Kuno:1999jp}
\beq
\left.\frac{dN_{\gamma}}{dE'}\right|_{\mu,\,rad} &=&
\lrf{2}{m_{\mu}}\frac{\alpha}{36\pi}\lrf{1-x}{x}\Theta(1-x-r)
\label{eq:murad}
\\ 
&&
{}\hspace{-1.5cm}
\times\Bigg(
12\big[3-2x(1-x)^2\big]\ln\!\lrf{1-x}{r} + x(1-x)(46-55x)  - 102
\Bigg) \ ,
\nnmb
\eeq
where $x = 2E'/m_{\mu}$ and $r = (m_e/m_{\mu})^2$.
The radiative photon spectrum in the lab frame is then obtained 
by boosting as in Eq.~\eqref{eq:muboost}.

\subsubsection*{$\mathbf{V\to \pi^+\pi^-}$}
\vspace{-0.3cm}
Vector decays in this channel produce electrons primarily through the chain
$\pi^{-} \to \bar{\nu}_{\mu}\mu^-$ with $\mu^-\to \nu_{\mu}\bar{\nu}_ee^-$
(and its conjugate).  We focus on this chain exclusively since the
inclusive branching $\text{BR}(\pi^+\to e^+{\nu}_e) \simeq 1.2344\times 10^{-4}$~\cite{Aguilar-Arevalo:2015cdf}
is small enough to be numerically insignificant in our analysis.
Photons are also produced through FSR and intermediate radiative decays.  

  To obtain the leading-order electron spectrum, we focus on the $\pi^-$ branch
with $\pi^-\to \bar{\nu}_{\mu}\mu^-$
since an identical contribution arises for the $\pi^+$ branch.
In the $\pi^-$ rest frame, the $\mu^-$ is produced with Lorentz factor
\beq
\gamma' = \lrf{m_{\pi}}{2m_{\mu}}\left(1+\frac{m_{\mu}^2}{m_{\pi}^2}\right) \ .
\eeq
Aligning the $z$-axis with the direction of the outgoing muon,
the muon obtains polarization $P=+1$.  
Defining $E''$ as the electron energy in the muon rest frame
and $\theta''$ as the electron direction relative to the $z$-axis,
the energy spectrum in this frame is given by Eq.~\eqref{eq:michel}
(with $E'\to E''$ and $\theta'\to \theta''$).
The electron energy $E'$ spectrum in the pion rest frame then follows from
the boosting method described above with 
\beq
E' = \gamma'(E''+\beta'p''\cos\theta'')\ ,~~~
p_z' = \gamma'(p''\cos\theta'' + \beta' E'') \ ,~~~
p_x' = p''_x \ ,~~~
p_y' = p''_y \ .
\eeq
One more boost in an arbitrary pion direction 
$\hat{n}=(\sin\alpha\,\cos\phi,\sin\alpha\sin\phi,\cos\alpha)$ 
with Lorentz factor $\gamma = m_V/2m_{\pi} = 1/\sqrt{1-\beta^2}$ 
is needed to transform to the lab frame where the vector is at rest.  
The electron energy in this frame is
\beq
E = \gamma\,E'\left[1+\beta(\sin\alpha\cos\phi\sin\theta'
+\cos\alpha\,\cos\theta')\right] \ ,
\eeq
with 
\beq
\cos\theta' = \frac{\cos\theta''+\beta}{1+\beta\cos\theta''} \ .
\eeq
The lab frame electron energy distribution is then obtained by
a simple change of variables using uniform distributions for
$\cos\theta''\in [-1,1]$, $\cos\alpha \in [-1,1]$, $\phi \in [0,2\pi]$,
and $E''$ given by the Michel spectrum of Eq.~\eqref{eq:michel}.
Note that for the positron spectrum from antimuon decays, 
the positron polarization is $P=-1$ while the sign of the $P$ term
in the Michel spectrum is reversed leading to the same distribution
in $E''$ and $\cos\theta''$.

  The photon spectrum from this vector decay channel is obtained from
a combination of FSR and internal radiative decays as in Eq.~\eqref{eq:mugam}.  
For the FSR part,
from the $V \to \pi^+\pi^-$ step, we use the result 
of Ref.~\cite{Coogan:2019qpu},
\beq
\left.\frac{dN}{dE_\gamma}\right|_{\text{FSR}} &=& 
\lrf{2}{m_V}\!\frac{2\alpha}{\pi}\,
\frac{1}{(1-4 \mu _\pi ^2)^{3/2}}\; \frac{1}{x}\; \Bigg( 
\label{eq:fsrscl}
\\ 
&& 
{}\hspace{0.5cm}
(1-4\mu_{\pi}^2 ) (1-x-2\mu_{\pi}^2) \ln\!\lrf{1+ {\cal W}_\pi }{1- {\cal W}_\pi } 
\nnmb\\ 
&&
{}\hspace{2.3cm}
- [1-x-x^2-4 \mu _\pi ^2 (1-x)]{\cal W}_\pi \Bigg) \ ,
\nnmb
\eeq
where $x = E_{\gamma}/(m_V/2)$, $\mu_\pi = m_{\pi}/m_V$, 
and $\mathcal{W}_\pi = \sqrt{1-4\mu_\pi^2/(1/x)}$.
The radiative contributions come from the $\pi^-\to \mu^-\bar{\nu}_{\mu}$
and $\mu^-\to e^-\bar{\nu}_e\nu_{\mu}$ steps of the decay chain.
These were studied along with the radiative contribution from
$\pi^-\to e^-\bar{\nu}_e$ in Ref.~\cite{Coogan:2019qpu} where it was
found that the total contribution is nearly completely dominated
by the muon decay step. The contribution to the photon spectrum
then follows by applying the boosting procedure described above
to the photon spectrum of Eq.~\eqref{eq:murad}.

\subsubsection*{$\mathbf{V\to \pi^0\gamma}$}
\vspace{-0.3cm}
This channel produces a pair of boosted photons from the $\pi^0$ decay
as well as a monochromatic photon directly.  The $\pi^0$ and monochromatic 
photon energies in the lab frame are
\beq
E_0 \ = \ \frac{m_V}{2}\left(1+\frac{m_{\pi^0}^2}{m_V^2}\right) \ ,~~~~~
E_\gamma \ = \ \frac{m_V}{2}\left(1-\frac{m_{\pi^0}^2}{m_V^2}\right) \ .
\eeq
For the photons from the $\pi^0$ decay, we have
$E'=m_{\pi^0}/2$ in the $\pi^0$ rest frame and a Lorentz factor
$\gamma = E_0/{m_{\pi^0}} = 1/\sqrt{1-\beta^2}$. 
Summing these contributions and applying our previous results,
the full photon spectrum is therefore
\beq
\frac{dN_{\gamma}}{dE} \ = \ \delta(E-E_{\gamma}) 
+ \frac{2}{\beta\gamma\,m_{\pi^0}}\,B(E_0) \ ,
\eeq
where
\beq
B(E_0) &=& 
\Theta(E-(1-\beta)E_0/2)\;\Theta((1+\beta)E_0/2-E)\nnmb\\
&& 
\label{eq:bfunc}\\
&=&
\left\{
\begin{array}{ccl}
1&;&E ~\in~ [(1-\beta),\,(1+\beta)]\times (E_0/2)\\
0&;&\text{otherwise}
\end{array}
\right.\nnmb 
\eeq

\subsubsection*{$\mathbf{V\to \pi^0\pi^+\pi^-}$}
\vspace{-0.3cm}
For this channel we concentrate exclusively on the direct photons and
electrons that arise from the dominant $\pi^0\to \gamma\gamma$
and $\pi^-\to \mu^-\bar{\nu}_\mu$, $\mu^-\to e^-\bar{\nu}_e\nu_{\mu}$ decay chains.
The charged and neutral pions are created with a distribution of 
energies of
\beq
p_{3\pi}(E_+,E_-) \ = \
\frac{1}{\Gamma_{3\pi}}\,\frac{d^2\Gamma_{3\pi}}{dE_+dE_-} \ ,
\eeq
with $E_0 = m_V - E_+-E_-$, $\Gamma_{3\pi}$ given by Eq.~\eqref{eq:3pi},
and $d^2\Gamma_{3\pi}/dE_+dE_-$ given by the same expression but
with the function $\mathcal{I}(m_V^2)$ replaced by the integrand
of Eq.~\eqref{eq:3piint}.  The photon distribution is thus just
the photon spectrum from a boosted $\pi^0$ decay found above
weighted by the distribution of $\pi^0$ energies,
\beq
\frac{dN_{\gamma}}{dE} \ = \ \int\!dE_+\int\!dE_-\;p_{3\pi}(E_+,E_-)\,
\frac{2}{\beta\gamma\,m_{\pi^0}}\,B(E_0) \ ,
\eeq
with $\gamma = 1/\sqrt{1-\beta^2} = E_0/m_{\pi^0}$
and $B(E_0)$ from Eq.~\eqref{eq:bfunc}.
Similarly, the electron (plus positron) spectrum is calculated using the 
boosted electron spectrum found previously for charged pion decay 
weighted by the joint distribution of $E_+$ and $E_-$ energies.

\section{BBN Bounds on Sub-GeV Energy Injection\label{sec:bbn}}

  Energy injected into the cosmological plasma can disrupt the predictions 
of standard BBN by altering the ratio of neutrons to protons 
or by destroying/creating light elements after they are formed. 
In this section we study the photodissociation of light elements 
from electromagnetic energy injected by decays of long-lived sub-GeV dark vectors 
in the early universe.  This is expected to be the most important effect
of dark vector decays on BBN for decay lifetimes $\tau_V > 10^4\,\text{s}$
and masses greater than a few MeV.  We apply our results to constrain the 
pre-decay abundances of such vectors.

\subsection{Methods for Calculating the Impact on BBN}

  Decays of sub-GeV dark vectors produce photons, electrons, and neutrinos,
both directly and through intermediate muons and pions.  To compute 
photodissociation effects from these decays, we make use of the branching 
fractions and the photon and electron energy spectra computed in the previous
section.  Note that for the cosmological times at which photodissociation
is effective, $t\gtrsim 10^4\,\text{s}$, the muons and pions injected
by sub-GeV vectors decay before they are slowed significantly by interactions
with the cosmological background~\cite{Kawasaki:2004qu,Pospelov:2010cw}.

  Electrons and photons injected into the dense cosmological plasma generally
interact with the plasma before reacting with light nuclei.  The scattering
of hard primaries off  the background plasma creates an electromagnetic~(EM) 
cascade that remains strongly suppressed until well after the main element 
creation stage of BBN has completed.  Since the formation of the EM cascade 
is fast relative to the rate of scattering with nuclei, 
the resulting photon cascade spectrum can be treated as the source 
for subsequent photodissociation reactions on nuclei.

  To compute the electromagnetic cascade spectra of photons and electrons 
(with positrons counted as electrons here), we follow the methods 
of Ref.~\cite{Forestell:2018txr} based on the earlier works 
of Refs.~\cite{Protheroe:1994dt,Kawasaki:1994sc}.
Defining the differential number density per unit energy of photons or
electrons in the cascade by
\beq
\naa \ = \ \frac{dn_a}{dE} \ , ~~~~~a=\gamma,\,e 
\eeq
the cascade spectra evolve according to
\beq
\frac{d\naa}{dt}(E) = - \Gamma_a(E)\naa(E) + \mathcal{S}_a(E)  \ ,
\label{eq:casc1}
\eeq
where $\Gamma_a$ is the net damping rate for species $a$ and energy $E$ 
and  $\mathcal{S}_a(E)$ is the injection rate from all sources at this energy.
The relevant damping and transfer reactions are generally fast relative 
to the Hubble rate and the effective photodissociation rates with light nuclei, 
and therefore the quasistatic limit of $d\naa/dt \to 0$ is a good 
approximation~\cite{Cyburt:2002uv}.\footnote{With this in mind we have also left out a Hubble dilution term in Eq.~\eqref{eq:casc1}.}
This gives
\beq
\naa(E) = \frac{\mathcal{S}_a(E)}{\Gamma_a(E)} \ ,
\label{eq:casc2}
\eeq
with both terms varying adiabatically as functions of time (or temperature).

The damping rate $\Gamma_a$ in Eq.~\eqref{eq:casc1} describes any reaction 
that transfers energy from species $a$ at energy $E$ to lower energies.  
The source terms in the electromagnetic cascade receive contributions 
from direct injection as well as from transfer reactions
moving energy from higher up in the cascade down to energy $E$.  Together,
these take the explicit forms
\beq
\mathcal{S}_a(E) = R\,\frac{dN_a}{dE}
+ \sum_b\int_{E}^{E_{X}}\!dE'\;K_{ab}(E,E')\,\nbb(E') \ ,
\eeq
where $R$ is the injection rate, $dN_a/dE$ is the primary energy spectrum
per injection, $E_X$ is the maximum energy in the cascade,
and $K_{ab}(E,E')$ is transfer kernel describing reactions
$b(E')+X_{BG}\to a(E)+X'_{BG}$ within the cascade. In the case of energy
injection from the decays of species $V$, 
$E_X \leq m_V/2$ and the injection rate is
\beq
R = \frac{n_V^0}{\tau_V}\,e^{-t/\tau_V} \ ,
\label{eq:rrate}
\eeq
where $\tau_V$ is the decay lifetime and $n_V^0$ is the number density 
the species would have in the in the absence of decays.

  Multiple reactions contribute to the damping and source terms in the cascade.
For photons, the dominant contribution to damping at higher energies 
$E\geq E_c \equiv m_e^2/22T$ is the photon-photon pair production~(4P) 
reaction, $\gamma+\gamma_{BG}\to e^++e^-$~\cite{Kawasaki:1994sc}. 
In the case of electrons, at the relevant energies damping is dominated 
by inverse Compton~(IC) scattering 
$e^{\mp}+\gamma_{BG} \to e^{\mp}+\gamma$~\cite{Blumenthal:1970gc}.  
Together, these processes strongly suppress the electromagnetic cascade 
below nuclear dissociation thresholds until after the period of 
element formation, with $E_c > 2\,\mev$
only for $T \lesssim 6\,\kev$~($t\gtrsim 7600\,\text{s}$). 
As a result, BBN bounds on electromagnetic injection from decay fall
off very quickly for lifetimes $\tau_V \lesssim 10^4\,\mathrm{s}$.

  For very high-energy initial injection, with $E_XT \gg m_e^2$,
the photon cascade spectrum resulting from the 4P, IC, and other reactions
is found to have a universal form.  Specifically, the universal photon spectrum
depends only on the total amount of electromagnetic energy injected into
the cosmological plasma, and not on the detailed injection spectra or
whether it came in the form or photons or electrons~\cite{Protheroe:1994dt,Kawasaki:1994sc}.  
However, recent studies of electromagnetic injection at lower energies 
have found significant deviations from universality~\cite{Forestell:2018txr,Poulin:2015woa,Poulin:2015opa,Hufnagel:2018bjp}. 
These deviations typically occur when the primary injection energy falls 
below the 4P cutoff $E_c$ or nuclear dissociation thresholds.
For the sub-GeV dark vectors of interest in this work, we find that this
takes place throughout a very significant portion of the relevant parameter space.
More details on deviations from universality and when it occurs are
collected in Ref.~\cite{Forestell:2018txr} and Appendix~\ref{sec:appb}.

With the photon cascade spectrum in hand, the effect of photodissociation
on the light element abundances can be described by Boltzmann equations
of the form
\begin{eqnarray}
\frac{dY_A}{dt} = \sum _i Y_i \int_0^\infty\!d\eg\,  
\ngam(\eg)\,\sigma _{\gamma +i \to A}(\eg) 
- Y_A \sum_f \int_0^{\infty }\!\;d\eg\,\ngam(\eg)\,\sigma_{\gamma + A \to f}(\eg) \ ,
\label{eq:bbnboltz}
\end{eqnarray}
where $\ngam(\eg)$ are the photon cascade spectra,
$A$ and the sums run over the relevant isotopes, 
and $Y_A$ are isotope number densities normalized to the entropy density,
\beq
Y_A = \frac{n_A}{s} \ .
\eeq
Reactions initiated by electrons are not included because their cascade
spectrum is always strongly suppressed by IC scattering.

The nuclear species included in our analysis are
hydrogen~(H), 
deuterium~($\mathrm{D} = {}^2\mathrm{H}$),
tritium~($\mathrm{T} = {}^3\mathrm{H}$),
helium-3~(${}^3\mathrm{He}$),
and helium~($\mathrm{He} = {}^4\mathrm{He}$).
Heavier species such as lithium have much smaller primordial abundances
and their inclusion would not alter our results for the light elements 
we consider.  For the nuclear cross sections in Eq.~\eqref{eq:bbnboltz}, 
we use the simple parametrizations collected in Ref.~\cite{Cyburt:2002uv}.  
These cross sections all have the same general shape with a sharp rise 
at threshold up to a peak followed by a smooth fall off.
The two most important thresholds for our study are $E_{th}\simeq 2.22\,\mev$
for deuterium photodissociation and $E_{th} \simeq 19.81\,\mev$ for helium, 
with the other relevant reaction thresholds typically in the range 
of $E_{th} \sim 2.5-8.5\,\mev$.

To solve the coupled evolution equations of Eq.~\eqref{eq:bbnboltz},
we convert time to redshift and use initial primordial abundances
expected from standard BBN predicted by \texttt{PArthENoPE}~\cite{Pisanti:2007hk,Consiglio:2017pot}:
\beq
Y_p = 0.247 \ ,~~~~~
\frac{n_{\mathrm{D}}}{n_\mathrm{H}} = 2.45\times 10^{-5} \ ,~~~~~
\frac{n_{^3\mathrm{He}}}{n_\mathrm{H}} = 0.998\times 10^{-5} \ .
\eeq
We then compare the resulting light element abundances to the following
observed values, quoted with effective $1\sigma$ uncertainties
(with theoretical and experimental uncertainties combined in quadrature):
\beq
Y_p &=& 0.245\pm 0.004 
\hspace{2.22cm}(\text{Ref.~\cite{Aver:2015iza}})\\
\frac{n_{\mathrm{D}}}{n_\mathrm{H}} &=& (2.53\pm 0.05)\times 10^{-5}
\hspace{1.0cm}(\text{Ref.~\cite{Cooke:2017cwo}}) \\
\frac{n_{^3\mathrm{He}}}{n_\mathrm{H}} &=& (1.0\pm 0.5)\times 10^{-5} 
\hspace{1.4cm}(\text{Ref.~\cite{Geiss:2003ab}})
\ .
\eeq
The helium mass fraction $Y_p$ we use is consistent 
with Ref.~\cite{Peimbert:2016bdg} and earlier determinations
but significantly lower than the result of Ref.~\cite{Izotov:2014fga}.
The uncertainty on the ratio $n_{\mathrm{D}}/{n_\mathrm{H}}$ is dominated
by a theory uncertainty on the rate of photon capture on deuterium
from Ref.~\cite{Marcucci:2015yla}.  For $n_{^3\mathrm{He}}/n_\mathrm{H}$,
we use the determination of $(n_{\mathrm{D}}+n_{^3\mathrm{He}})/n_{\mathrm{H}}$
of Ref.~\cite{Geiss:2003ab} along with the value
of $n_{\mathrm{D}}/{n_\mathrm{H}}$ from Ref.~\cite{Cooke:2017cwo}.  
The uncertainties quoted here are generous, and in the analysis below 
we implement exclusions at the 2$\sigma$ level.

\begin{figure}[ttt]
 \begin{center}
   \includegraphics[width = 0.33\textwidth]{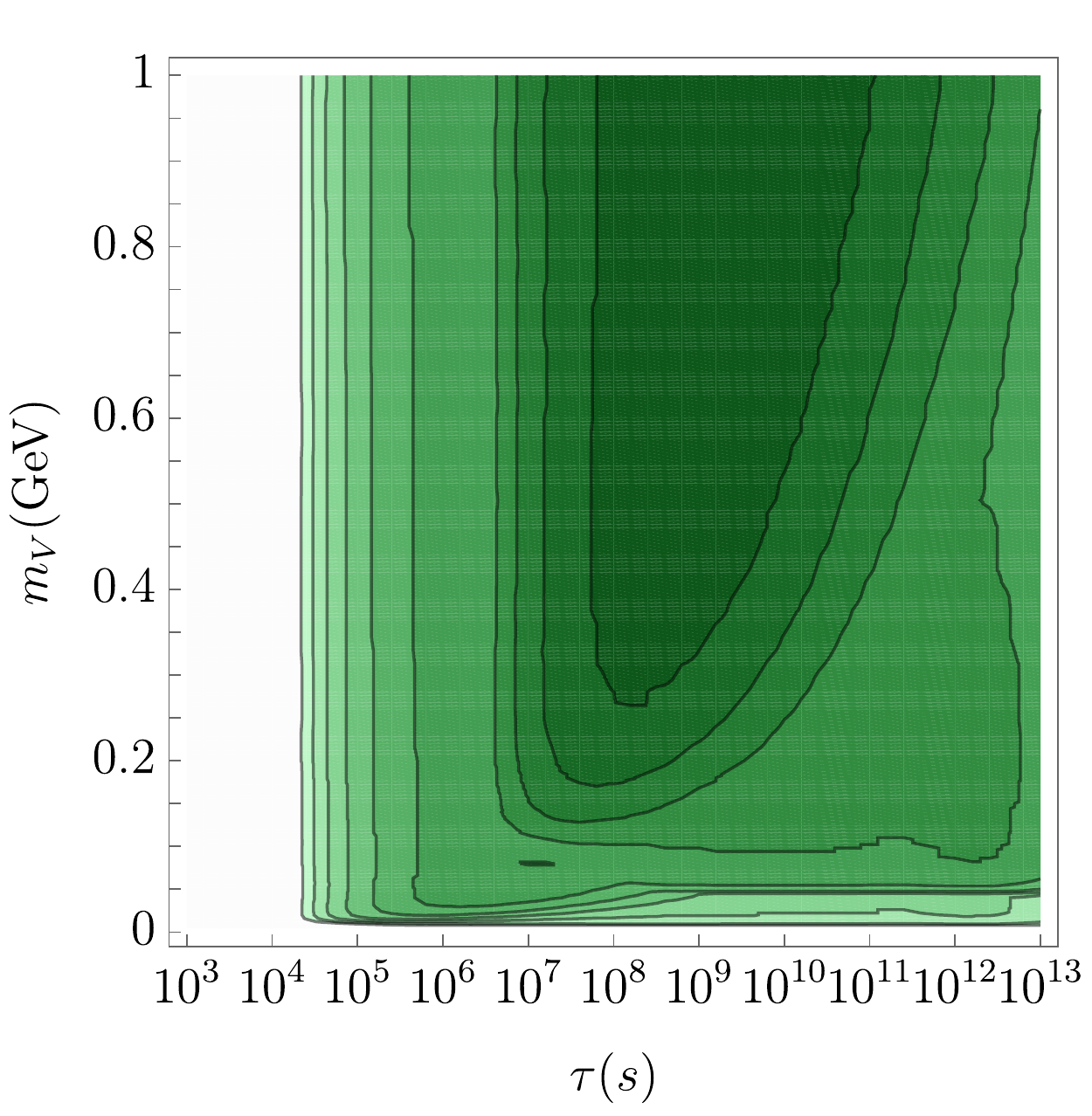}~~
   \includegraphics[width = 0.33\textwidth]{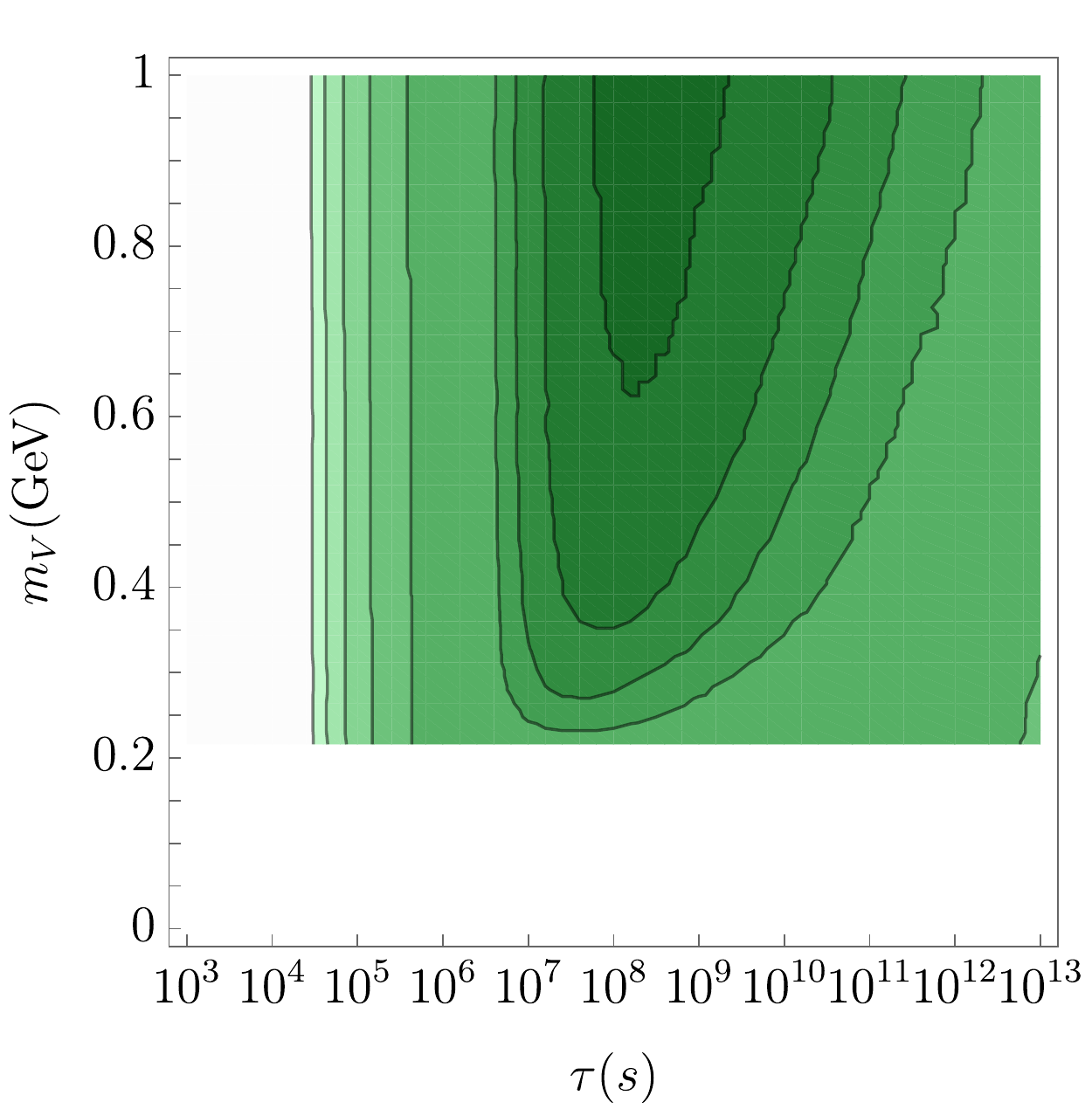}~~
   \hspace{1.55cm}
   \includegraphics[width = 0.077\textwidth]{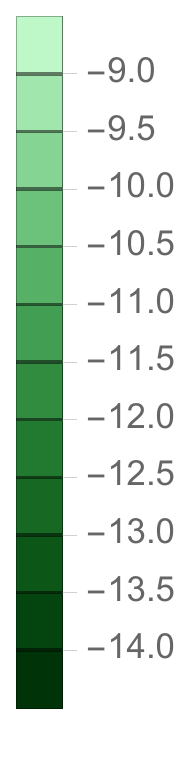}
   \hspace{1.5cm}
   \vspace{0.5cm}\\
   \includegraphics[width = 0.33\textwidth]{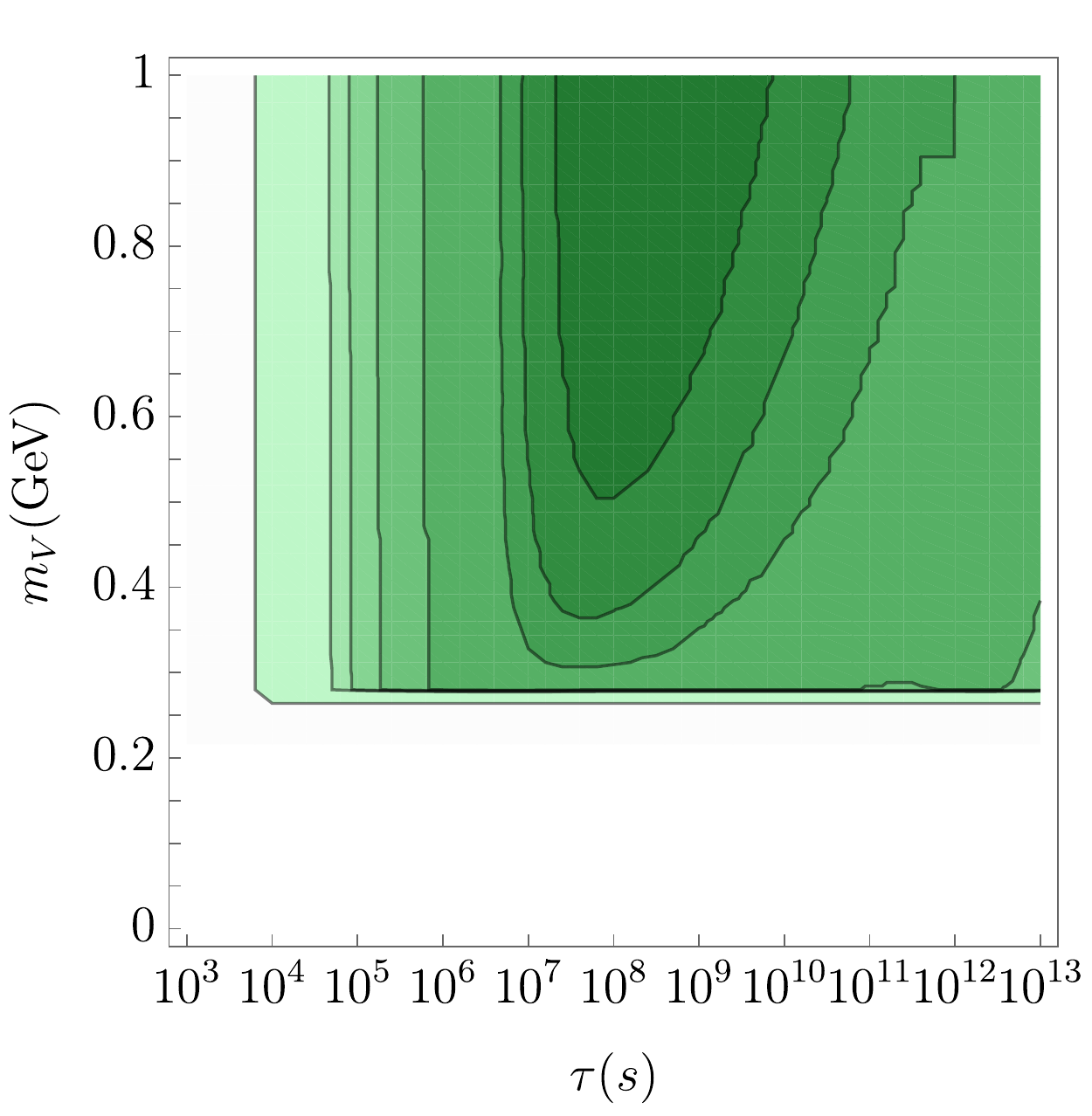}~~
   \includegraphics[width = 0.33\textwidth]{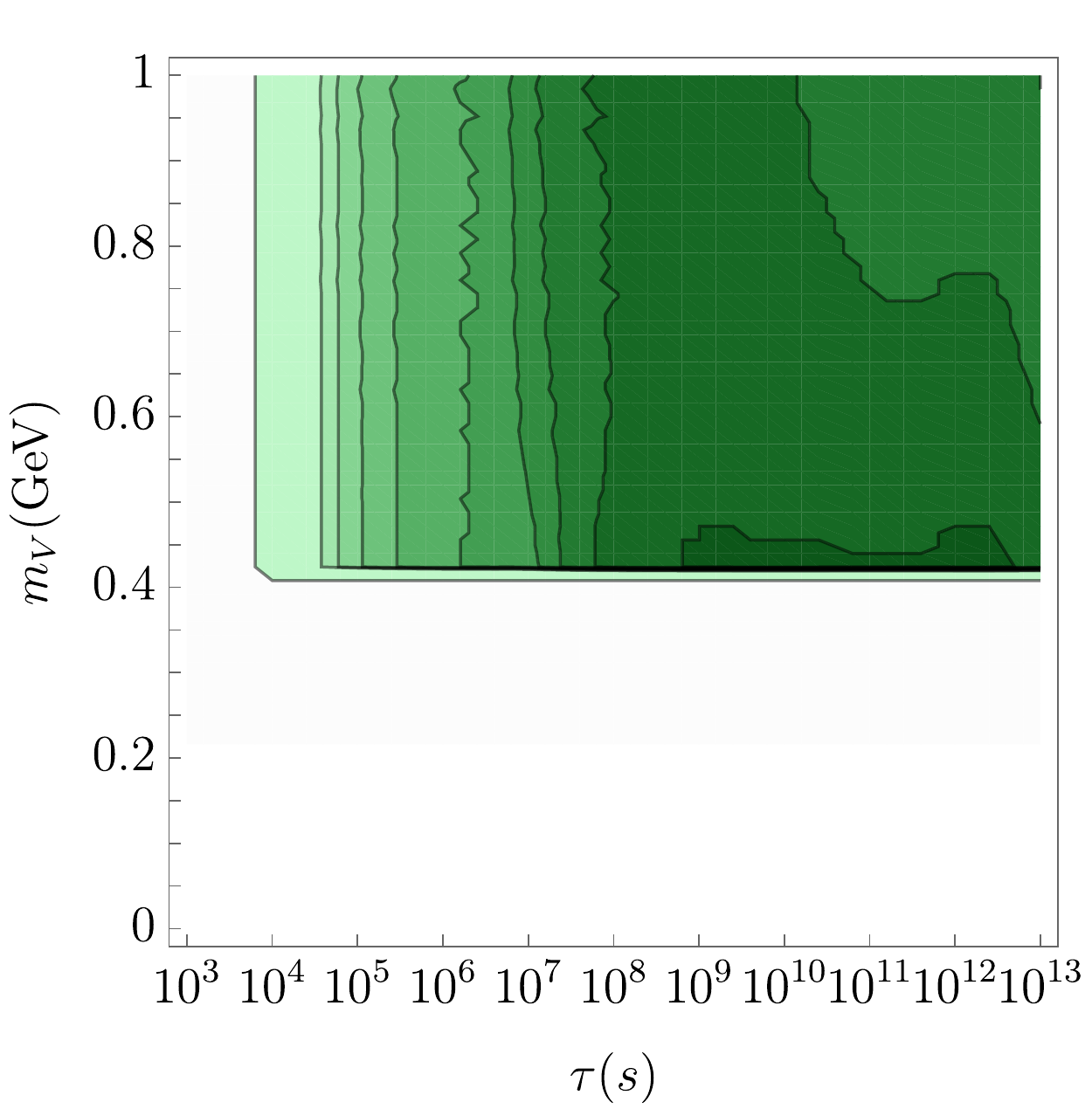}~~
   \includegraphics[width = 0.33\textwidth]{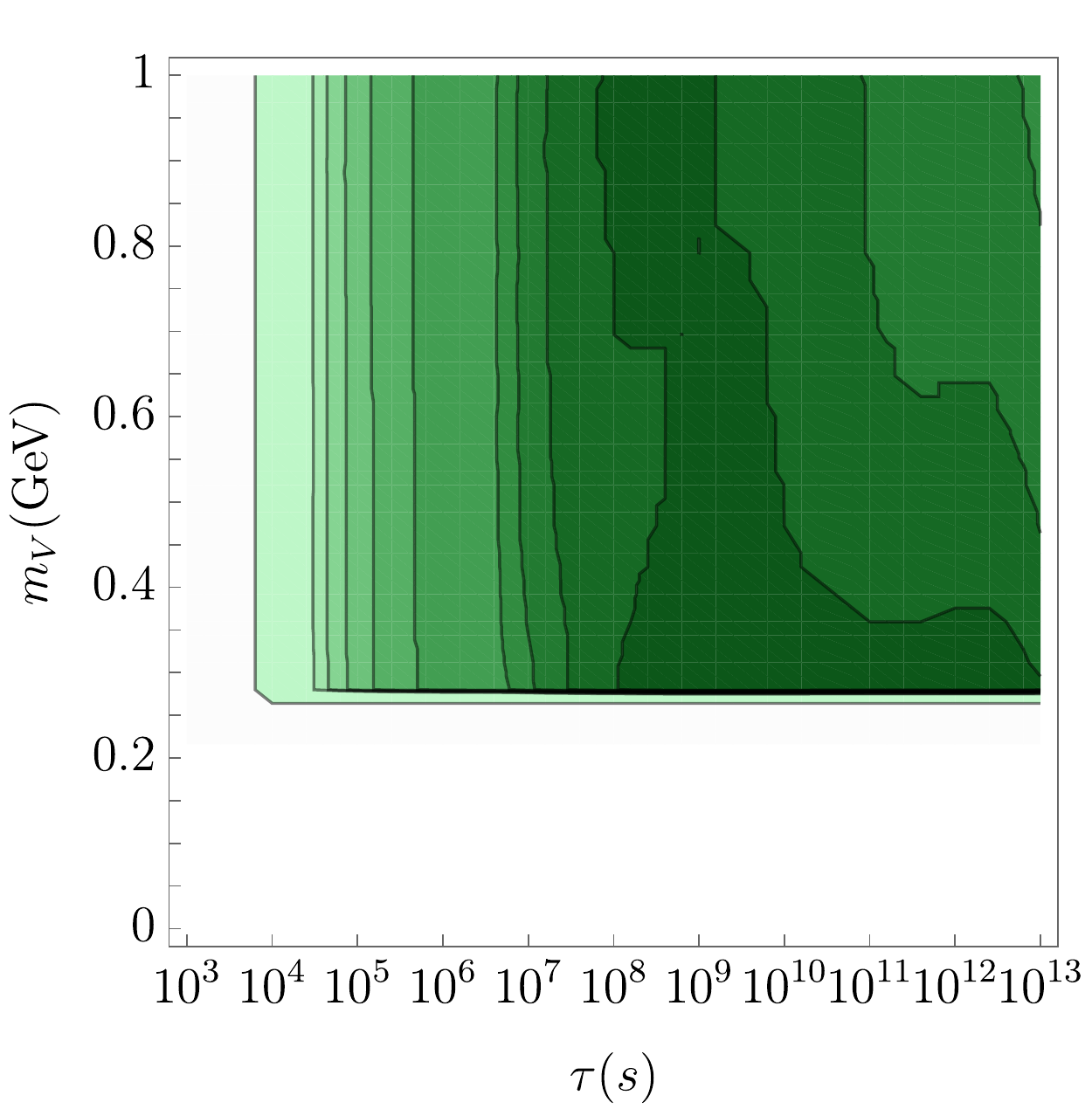}
 \end{center}
\vspace{-0.7cm}
 \caption{
BBN upper limits from electromagnetic effects on the mass times yield 
$\log_{10}(m_VY_V/\gev)$ of a sub-GeV decaying particle
$V$ in the lifetime-mass plane for the exclusive decays 
$V\to e^+e^-$~(upper left), 
$V\to \mu^+\mu^-$~(upper right), 
$V\to \pi^+\pi^-$~(lower left), 
$V\to \pi^+\pi^-\pi^0$~(lower middle), 
and
$V\to \pi^0\gamma$~(lower right).
}
 \label{fig:bbnex}
\end{figure}

  To illustrate our approach, we show in Fig.~\ref{fig:bbnex} the 
photodissociation bounds from BBN on the pre-decay yield $Y_V = n_V/s$ 
for a decaying species $V$ with mass $m_V$ and lifetime $\tau_V$ 
assuming that all decays occur exclusively via one of the channels
$V\to e^+e^-$, $\mu^+\mu^-$, $\pi^+\pi^-$, $\pi^0\pi^+\pi^-$, $\pi^0\gamma$.
Decays to $e^+e^-$ produce the strongest effect throughout most of the 
lifetime--mass plane shown.  Relative to the $\mu^+\mu^-$ and $\pi^+\pi^-$
channels, electron decays produce more energetic electrons and a greater
electromagnetic injection fraction.  At lower masses and larger lifetimes,
in the lower right region of the plots,
the injected electrons scatter off background photons in the Thomson
regime where the upscattered photons receive energies well below
nuclear photodissociation thresholds.  The dominant sources of photons
contributing to photodissociation are then FSR and radiative decays,
and these also tend to be greater for $e^+e^-$ than $\mu^+\mu^-$ 
or $\pi^+\pi^-$ channels.  This obstacle does not affect the 
$\pi^0\pi^+\pi^-$ and $\pi^0\gamma$ channels which produce photons
at the leading order and thus stronger BBN limits in the lower right
region of the plots.  Note that even after rescaling by the
total electromagnetic fractions produced by these various modes,
the resulting BBN limits are significantly different, 
reflecting the breaking of universality of the cascade photon spectrum, 
as discussed in more detail in Appendix~\ref{sec:appb}.

\subsection{BBN Bounds on Dark Vectors}

\begin{figure}[ttt]
 \begin{center}
   \includegraphics[width = 0.33\textwidth]{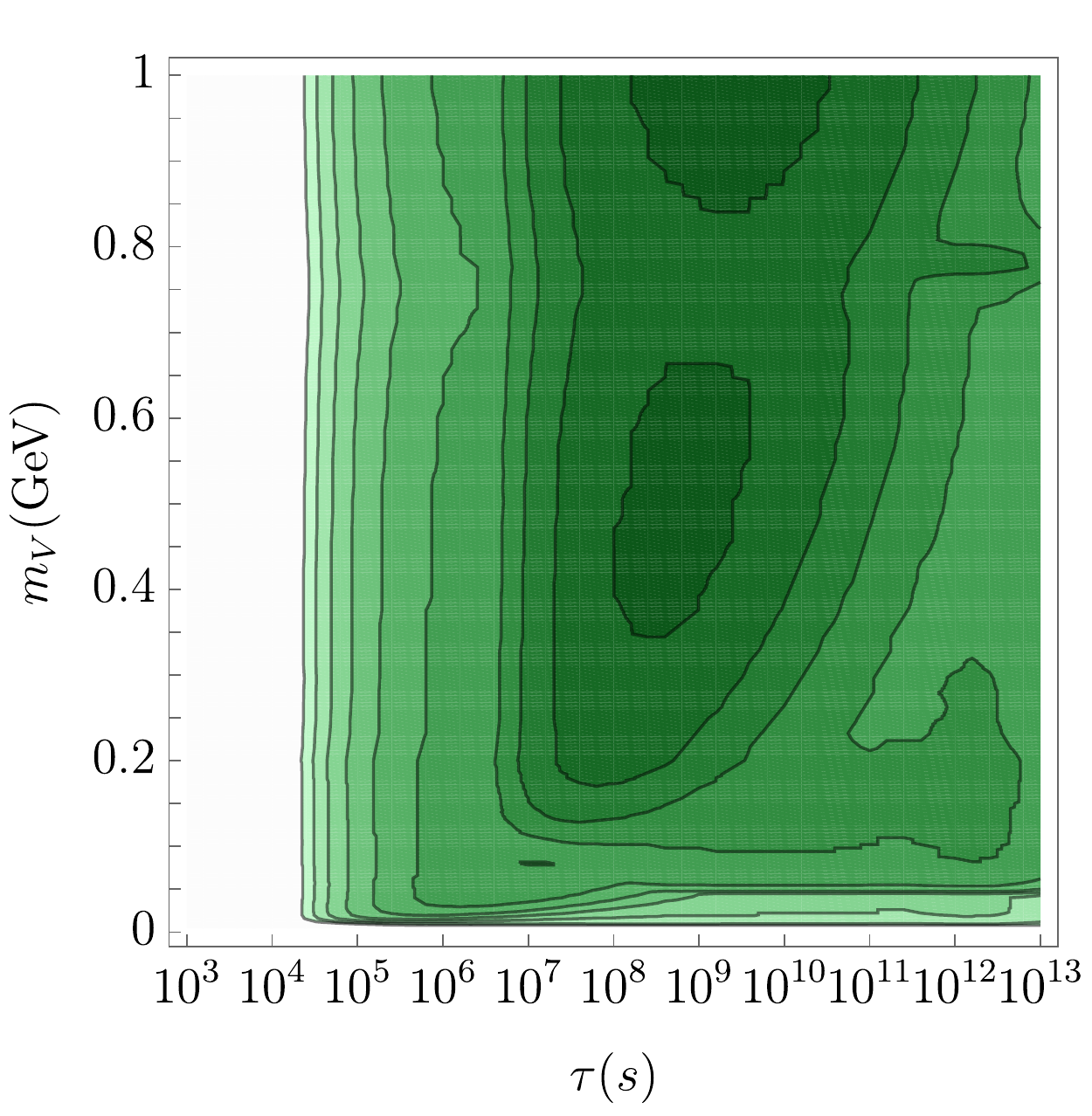}~~
   \includegraphics[width = 0.33\textwidth]{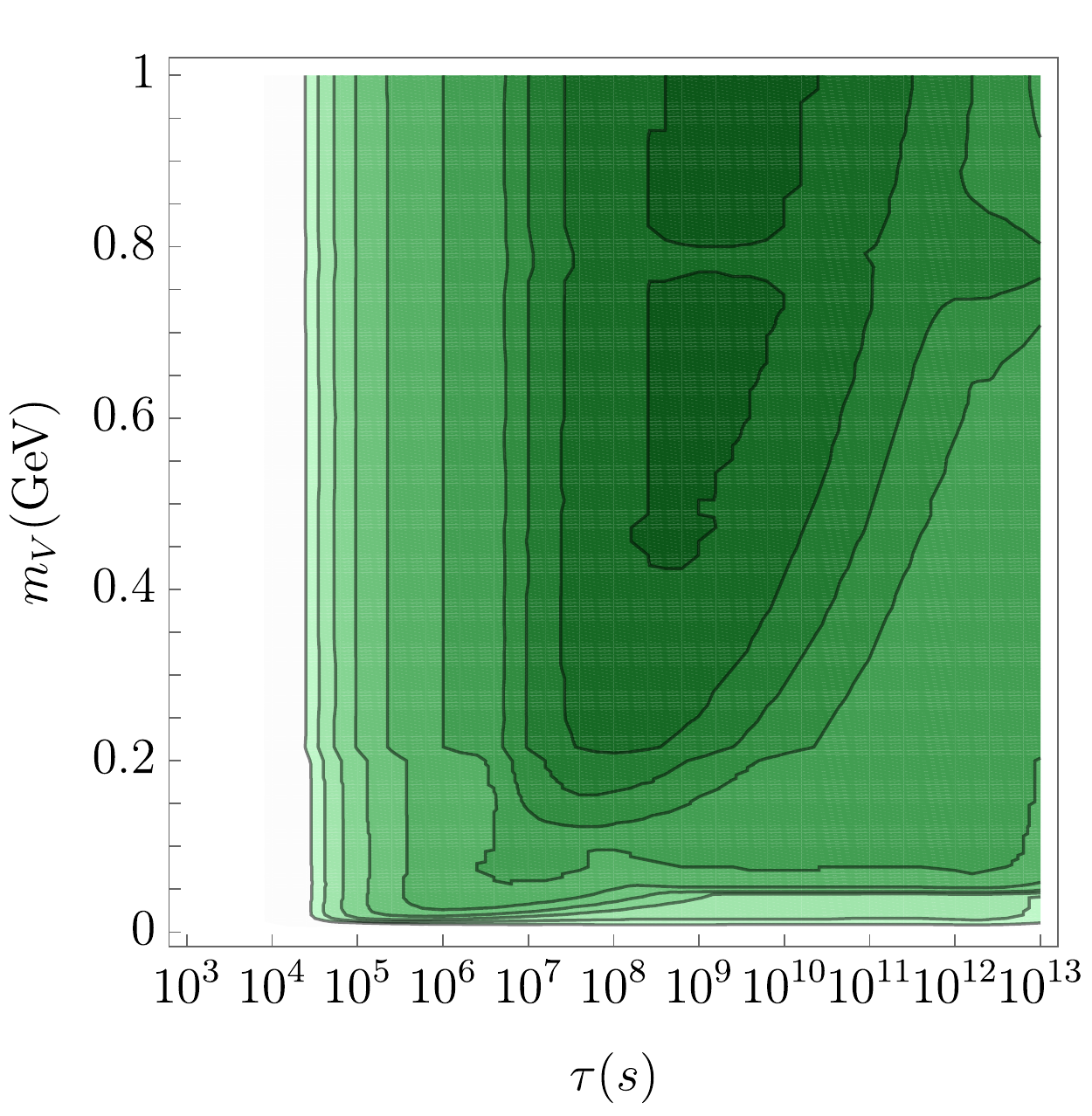}~~
   \hspace{1.55cm}
   \includegraphics[width = 0.077\textwidth]{bbn-legend.pdf}
   \hspace{1.5cm}
   \vspace{0.5cm}\\
   \includegraphics[width = 0.33\textwidth]{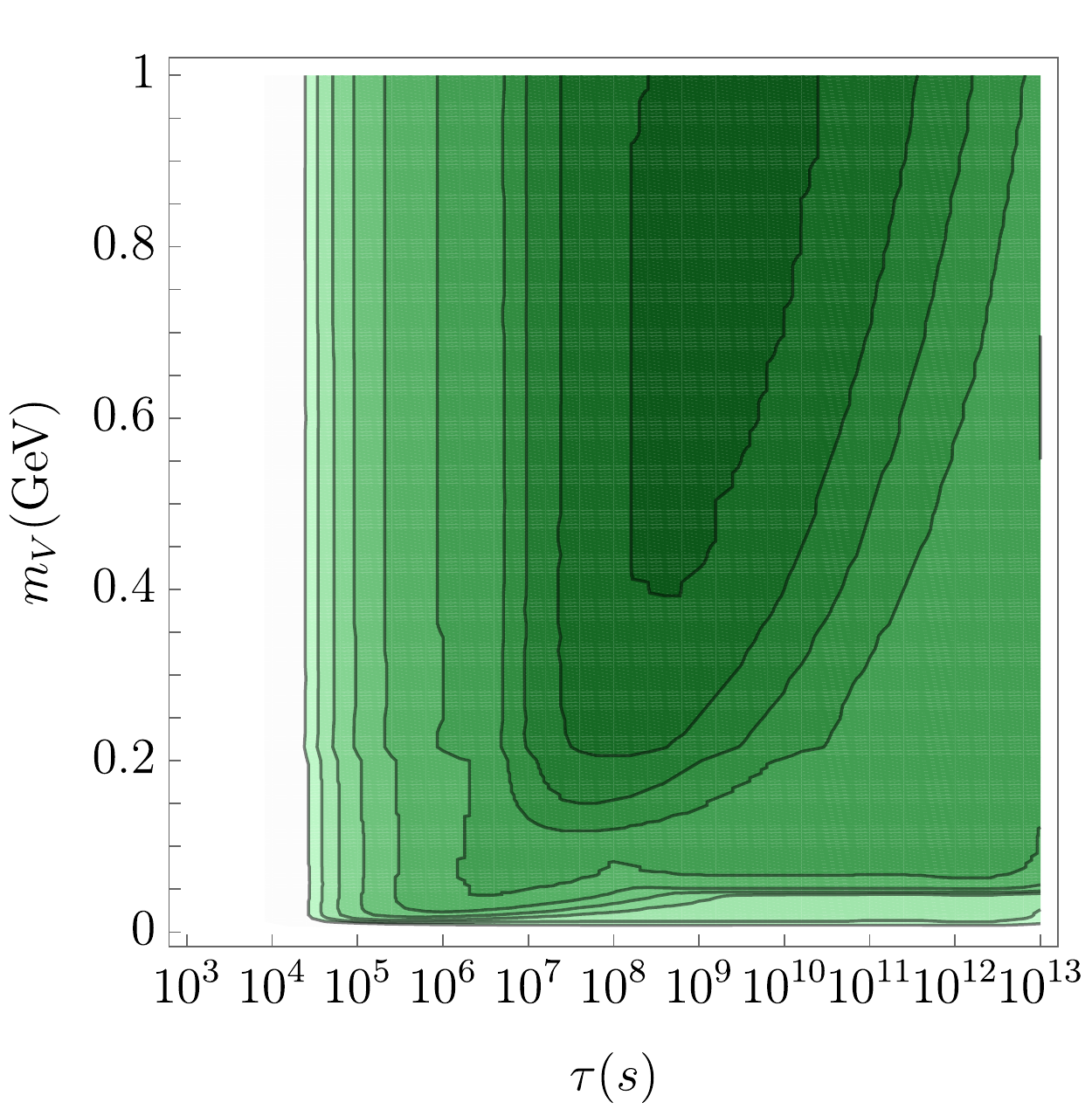}~~
   \includegraphics[width = 0.33\textwidth]{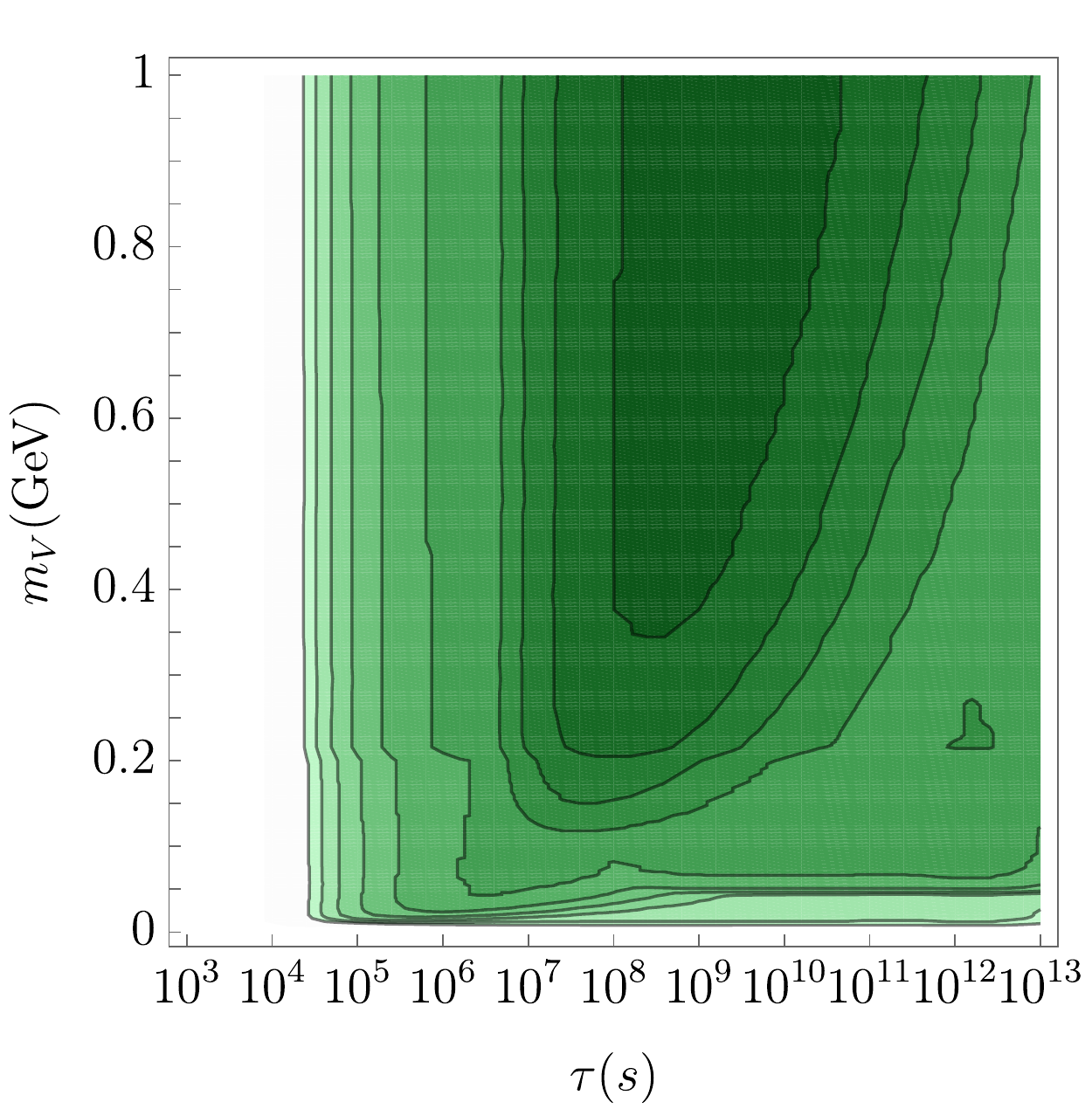}~~
   \includegraphics[width = 0.33\textwidth]{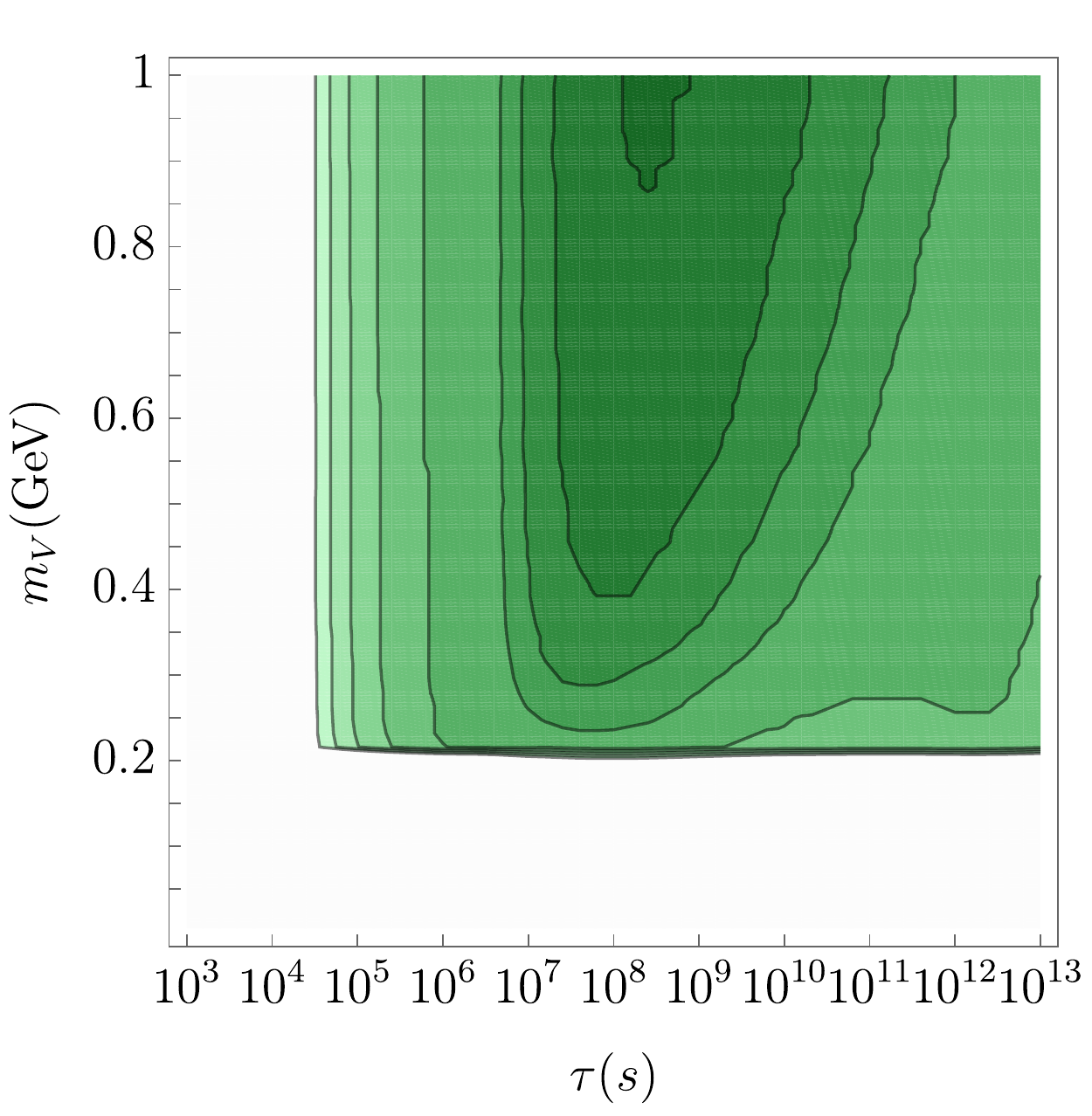}
 \end{center}
\vspace{-0.7cm}
 \caption{
BBN upper limits from electromagnetic effects on the mass times yield 
$\log_{10}(m_VY_V/\gev)$ of sub-GeV dark vectors 
in the lifetime-mass plane for the dark vector species
DP~(upper left),
$\vbl$~(upper middle),
$\vem$~(lower left),
$\vet$~(lower middle),
and
$\vmt$~(lower right).
}
 \label{fig:bbndp}
 \end{figure}

 Using the dark vector decay fractions and the BBN methods described
above, we can now derive BBN limits due to photodissociation
on the pre-decay yield $Y_V=n_V/s$
of a dark vector with mass $m_V$ and lifetime $\tau_V$.
Our results are shown in Fig.~\ref{fig:bbndp} for the dark vector species
DP~(upper left),
$\vbl$~(upper middle),
$\vem$~(lower left),
$\vet$~(lower middle),
and
$\vmt$~(lower right).
The shapes of these exclusions can be understood by comparing the
branching fractions shown in Fig.~\ref{fig:vprodbr} and the bounds
on the contributing decay modes in Fig.~\ref{fig:bbnex}.
For the dark photon~(DP) and $\vbl$ vectors, the bounds get weaker due to
hadronic contributions near the $\rho$ and $\omega$ resonances.
This effect is greater for the DP due to its strong mixing with the wide $\rho$
resonance relative to the $\vbl$ that approximately only mixes with 
the narrow $\omega$.  With the exception of the DP, an increase in the bound
is seen near the muon threshold at $m_V = 2m_\mu$ due to the 
larger visible mode decay branching fractions when the muon channel turns on
and a smaller fraction of the decays go to neutrinos.
For the $L_\mu\!-\!L_{\tau}$ vector, the bounds below the muon threshold
fall beneath the range covered by the figure since the decay products 
in this region are nearly completely dominated by neutrinos.

  In addition to photodissociation, dark vector decays can also modify the
outcome of standand BBN by altering the ratio of neutrons to protons 
and through the hadrodissociation of light elements.  While these effects
can be significant and further constrain the properties of dark vectors,
we argue that they are also largely orthogonal to the bounds from 
photodissociation derived above.   
Lighter dark vectors with $m_V \lesssim 20\,\mev$ can alter the 
effective number of light degrees of freedom during and after neutron-proton 
freeze-out through their direct influence~\cite{Berlin:2019pbq}, 
from their decays following neutrino decoupling~\cite{Berlin:2019pbq,Ibe:2019gpv},
or by equilibrating light right-handed neutrinos (which we assume to not 
be present)~\cite{Heeck:2014zfa,Barger:2003zh}.  The couplings required
for these effects to be significant correspond to dark vector lifetimes
below $\tau_V \lesssim 10^4\,\text{s}$, or large initial densities
$m_VY_V \gg 10^{-9}$, and thus these considerations apply to 
earlier decays or produce weaker constraints than photodissociation.
Heavier dark vectors that decay to pions can alter the neutron to proton ratio
through reactions such as $p+\pi^-\to n+\pi^0$ or destroy light
elements through hadrodissociation.  
The analyses of Refs.~\cite{Fradette:2014sza,Berger:2016vxi,Pospelov:2010cw}
find that these effects are strongest for lifetimes $\tau_V \sim 10^3\,\mathrm{s}$
and $m_VY_V \gtrsim 10^{-11}\,\gev$.  Comparing to our results for 
photodissociation, these two sets of bounds largely apply independently
of one another, with only a very small region of possible interference near
$\tau_V \sim 10^4\,\text{s}$. Taken together, these considerations justify
our earlier treatment of photodissociation bounds from BBN in isolation.

\section{CMB Bounds on Sub-GeV Energy Injection\label{sec:cmb}}

  Energy injection in the early universe can also modify the 
power and freqency spectra of the CMB.  In this section we describe
the methods that we use to compute these effects and show the constraints
they imply for dark vector decays.

\subsection{Methods for the CMB Power Spectrum}

  Energy injected during and after recombination can ionize
newly-formed atoms and broaden the surface of last scattering.  
These effects can modify the power spectra of CMB fluctuations
relative to the standard recombination history~\cite{Chen:2003gz}.  
Precision measurements of the CMB power spectra can therefore be used 
to constrain new sources of energy injection 
in this period~\cite{Padmanabhan:2005es,Zhang:2007zzh}.

  The total rate of energy injection per unit volume from a decaying species 
of mass $m_V$ is simply $m_VR$, where $R$ is the decay rate per unit
volume given in Eq.~\eqref{eq:rrate}.  For decays near or after recombination, 
there may be a delay between the initial decay and when the resulting energy 
is deposited within the cosmological medium~\cite{Slatyer:2009yq}.  
We treat this as in Refs.~\cite{Slatyer:2012yq,Poulin:2016anj} 
and write the total energy deposition rate per unit volume as
\beq
\lrf{d\rho}{dt}_{\!\!dep} = f(z)\,\frac{m_Vn_V^0}{\tau_V} \ ,
\eeq
where $n_V^0$ is the number density in the absence of decays and
\beq
f(z) = \frac{H(z) \sum _{a}
\int_{\ln(1+z)}^\infty\!\frac{d\ln(1+z') }{H(z')}
\int\!dE\; 
T^{(a)}(z',z,E)\,E\,\frac{dN_a}{dE}\,e^{-t(z')/\tau_V} }
{\sum _{b} \int\!dE\; E\,\frac{dN_b}{dE} } \ .
\eeq
Here, $dN_a/dE$ are the energy spectra per decay, with the sum in
the numerator running over $a=e,\gamma$ and
the sum in the denominator over $b=e,\gamma,\nu$.
The $T^{(a)}(z',z,E)$ are transfer functions that describe 
the fraction of energy deposited at $z$ for injection at energy $E$ 
and redshift $z'\leq z$.  Note that the exponential depletion due to decays has 
been incorporated into the deposition function $f(z)$ as 
in Ref.~\cite{Poulin:2016anj}.

  In addition to the total efficiency of energy deposition described by $f(z)$,
similar functions $f_i(z)$ and corresponding transfer functions $T_i^{{(a)}}$
can be defined for energy deposition into specific channels such 
as hydrogen ionization, helium ionization, heating of the cosmological medium, 
and photons with energies below $10.2\,\text{eV}$ that 
free-stream~\cite{Slatyer:2015kla}.  
Arrays of these transfer functions are collected in Ref.~\cite{nebel:2012}
and we apply them to our analysis.  
The effect of energy injection on the CMB power spectra is almost entirely
due to the ionization of hydrogen, with much smaller contributions
from helium ionization and Lyman-$\alpha$ photons~\cite{Galli:2013dna}.
As in Ref.~\cite{Slatyer:2015kla}, we define an ionization fraction 
$\chi_{ion}$ by
\beq
\chi_{ion}(z) = f_{ion}(z)/f(z) \ ,
\eeq
where $f_{ion}$ is the total efficiency of depositing energy into the
ionization of hydrogen and helium.

  A powerful method to connect the energy deposition function $f(z)$ to
bounds from CMB measurements was derived in Ref.~\cite{Finkbeiner:2011dx} 
based on a principal component method.  In this approach, orthogonal 
eigenfunctions with respect to the space of energy deposition histories 
are obtained from data (or data projections) that characterize the impact 
of energy injection on the CMB power spectra, with a marginalization 
over $\lcdm$ parameters included.  The significance of the energy
injection is then given by the orthogonal sum over the expansion coefficients
of $f(z)$ with respect to the eigenfunction basis weighted by their
corresponding eigenvalues.  Ordering the eigenfunctions by the sizes
of their eigenvalues, only a small set of these \emph{principal components}
need to be included to get an excellent approximation 
of the full result~\cite{Finkbeiner:2011dx}
(which requires a intensive computation with tools such 
as \texttt{CLASS}~\cite{Lesgourgues:2011re} 
and \texttt{CosmoMC}~\cite{Lewis:2002ah} 
for each model of interest).

To estimate the CMB bounds from Planck~\cite{Aghanim:2018eyx} 
on dark vector decays, we make use of principal component functions 
and eigenvectors over the space of energy deposition histories described 
in Refs.~\cite{Slatyer:2012yq,Finkbeiner:2011dx,Cline:2013fm} 
and collected in Ref.~\cite{nebel:2012}.  These were computed 
in Ref.~\cite{Finkbeiner:2011dx} using the ionization fraction 
$\chi_{base}(z)$ formulated in Refs.~\cite{Chen:2003gz,1985ApJ...298..268S}.
This was subsequently improved in Ref.~\cite{Slatyer:2015kla}
to account for losses to photons with energies below $10.2\,\text{eV}$
that do not contribute to ionization.  
To implement this improvement, we follow the prescription 
of Refs.~\cite{Madhavacheril:2013cna,Slatyer:2015jla,Slatyer:2016qyl}
by replacing $f(z)$ in the analysis with
\beq
\tilde{f}(z) \ = \ {\chi_c(z)}f(z)\Big/ {\chi_{base}(z)} \ .
\eeq
With this prescription, we estimate the bounds from Planck
and make projections for a cosmic variance limited~(CVL) experiment
with multipoles up to $\ell = 2500$
using the principal component functions of Ref.~\cite{nebel:2012}.
In the case of Planck, these functions were derived based on a projection 
of the experimental sensitivity 
rather than data~\cite{nebel:2012,Finkbeiner:2011dx}.  
However, this projection turns out to be remarkably accurate at predicting
the limits on $s$-wave dark matter annihilation derived from the 
Planck 2015~\cite{Ade:2015xua} and Planck 2018~\cite{Aghanim:2018eyx} data sets.
For decays, we also find that the projected limits for longer lifetimes
$\tau_V \gg 10^{13}\,\text{s}$ agree well with the principal components
derived for dark matter decay with respect to the space of 
energy injection spectra based on Planck 2015 data~\cite{Slatyer:2016qyl}.

\subsection{Methods for the CMB Frequency Spectrum}

Late-time energy injection can also distort the CMB frequency 
spectrum~\cite{Hu:1992dc,Hu:1993gc} from the nearly perfect blackbody
that is observed~\cite{Fixsen:1996nj}.  For decays after 
the decoupling of double-Compton scattering at redshift
$z_{th} \simeq 2.0\times 10^6$
but before the decoupling of Compton scattering 
at $z_C \simeq 5.2\times 10^4$,
the decay products equilibrate kinetically but generate 
an effective photon chemical potential~\cite{Hu:1992dc,Hu:1993gc}.
Decays after Compton decoupling but before recombination at $z_{rec}\simeq 1090$
produce a distortion that can be described by the Compton $y$ 
parameter~\cite{Hu:1992dc,Hu:1993gc}.
The current limits on $\mu$ and $y$ from COBE/FIRAS are~\cite{Fixsen:1996nj}
\beq
\mu < 9\times 10^{-5},~~~~~|y| < 1.5\times 10^{-5} \ ,
\eeq
while the proposed PIXIE satellite is projected to have sensitivity 
to constrain~\cite{Kogut:2011xw}
\beq
\mu < 1\times 10^{-8},~~~~~|y| < 2\times 10^{-9} \ .
\eeq
These limits can be used to constrain decays in the early universe.

An accurate expression for both types of distortions
is~\cite{Chluba:2011hw,Chluba:2013vsa,Chluba:2013pya,Chluba:2016bvg}
\beq
\frac{\mu}{1.401} &\simeq& \int\!dt\;
\mathcal{J}_{\mu}\lrf{\Delta\dot{\rho}_{\gamma}}{\rho_{\gamma}} \\
\frac{y}{1/4}~ &\simeq& \int\!dt\;
\mathcal{J}_{y}\lrf{\Delta\dot{\rho}_{\gamma}}{\rho_{\gamma}}
\eeq
where the $\mathcal{J}_i$ are window functions, $\rho_{\gamma}$ is
the energy density of photons, and $\Delta\dot{\rho}_{\gamma}$
is the rate of electromagnetic energy injection per unit volume.
For the window functions, we use ``method~C'' of Ref.~\cite{Chluba:2016bvg}:
\beq
\mathcal{J}_\mu(z) &=& e^{-(z/z_{th})^{5/2}}\,
\left[1-\exp\left(-\Big[(1+z)/5.8\times 10^4\Big]^{1.88}\right)\right]\\
\mathcal{J}_y(z) &=& \left(1 + \Big[({1+z})/{6\times 10^4}\Big]\right)^{-1}
\Theta(z-z_{rec}) \ .
\eeq
The energy injection profile for a decaying dark vector is
$\Delta\dot{\rho}_{\gamma} = f_{em}m_V R$, 
with $R$ given by Eq.~\eqref{eq:rrate},
and $f_{em}$ the electromagnetic energy fraction of the decay.  
For sub-GeV dark vectors we approximate this fraction by
\beq
f_{em} \ \simeq \ \text{BR}(e^+e^-) 
+ \frac{1}{3}\,\text{BR}(\mu^+\mu^-)
+\frac{1}{4}\,\text{BR}(\pi^+\pi^-)
+ \frac{1}{2}\,\text{BR}(\pi^+\pi^-\pi^0) 
+ \text{BR}(\pi^0\gamma) \ .
\eeq
Note that in contrast to the BBN and CMB power spectrum bounds discussed
above, this constraint on decaying particles is largely insentive to the 
energy spectrum of the decays.

\subsection{CMB Bounds on Dark Vectors}

\begin{figure}[ttt]
 \begin{center}
   \includegraphics[width = 0.33\textwidth]{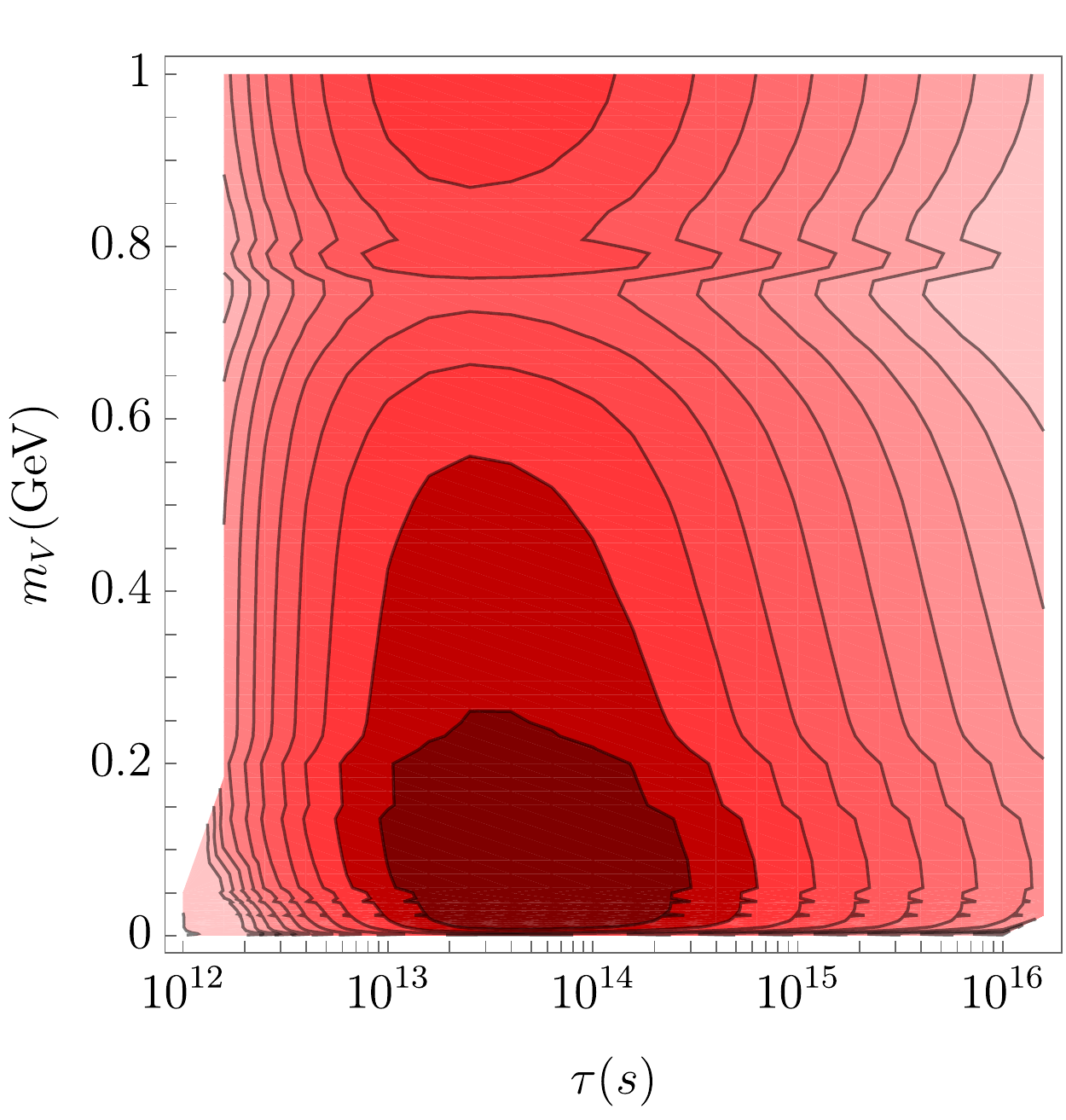}~~
   \includegraphics[width = 0.33\textwidth]{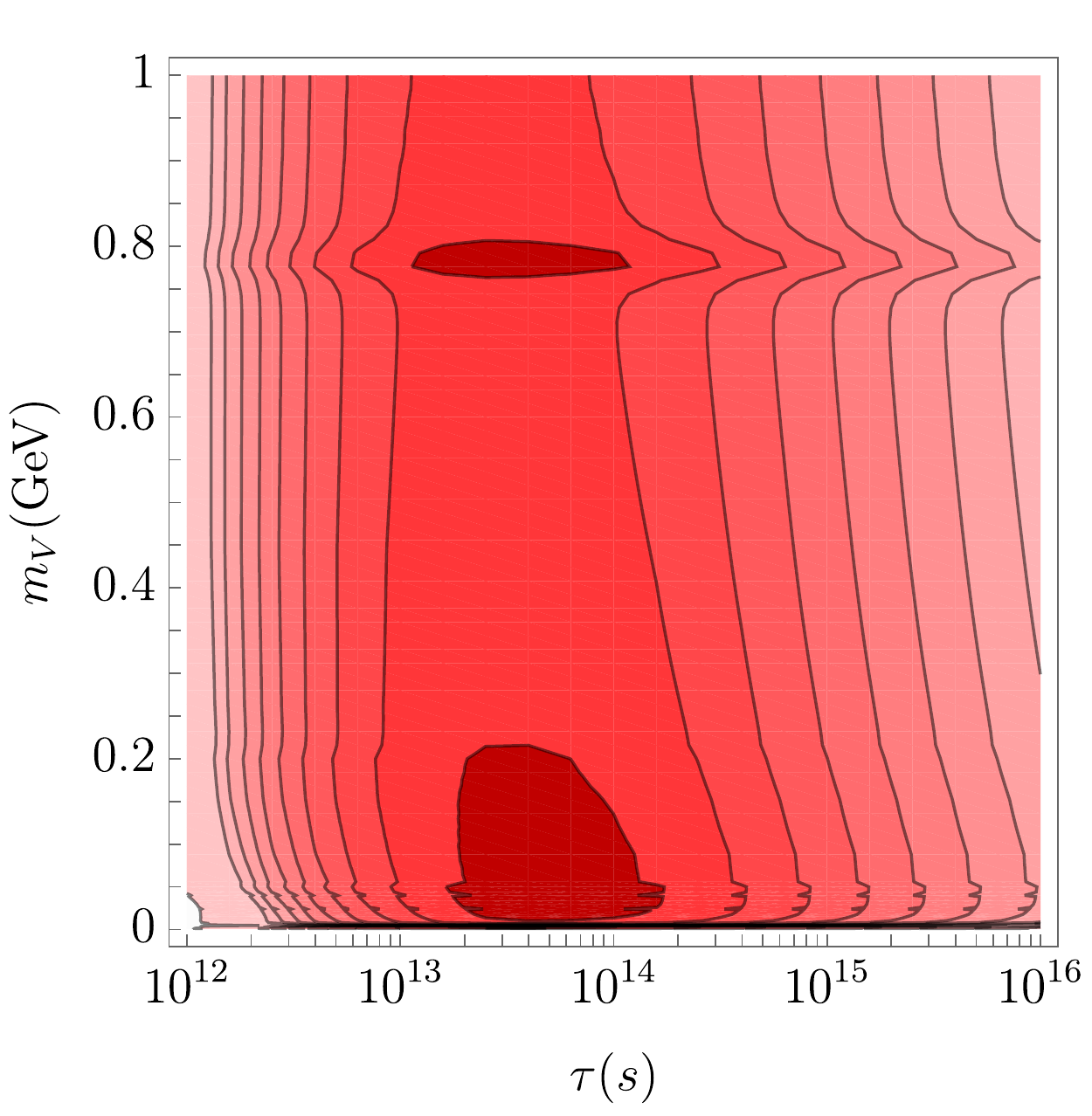}~~
   \hspace{1.55cm}
   \includegraphics[width = 0.077\textwidth]{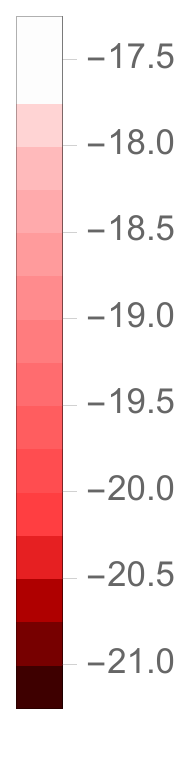}
   \hspace{1.5cm}
   \vspace{0.5cm}\\
   \includegraphics[width = 0.33\textwidth]{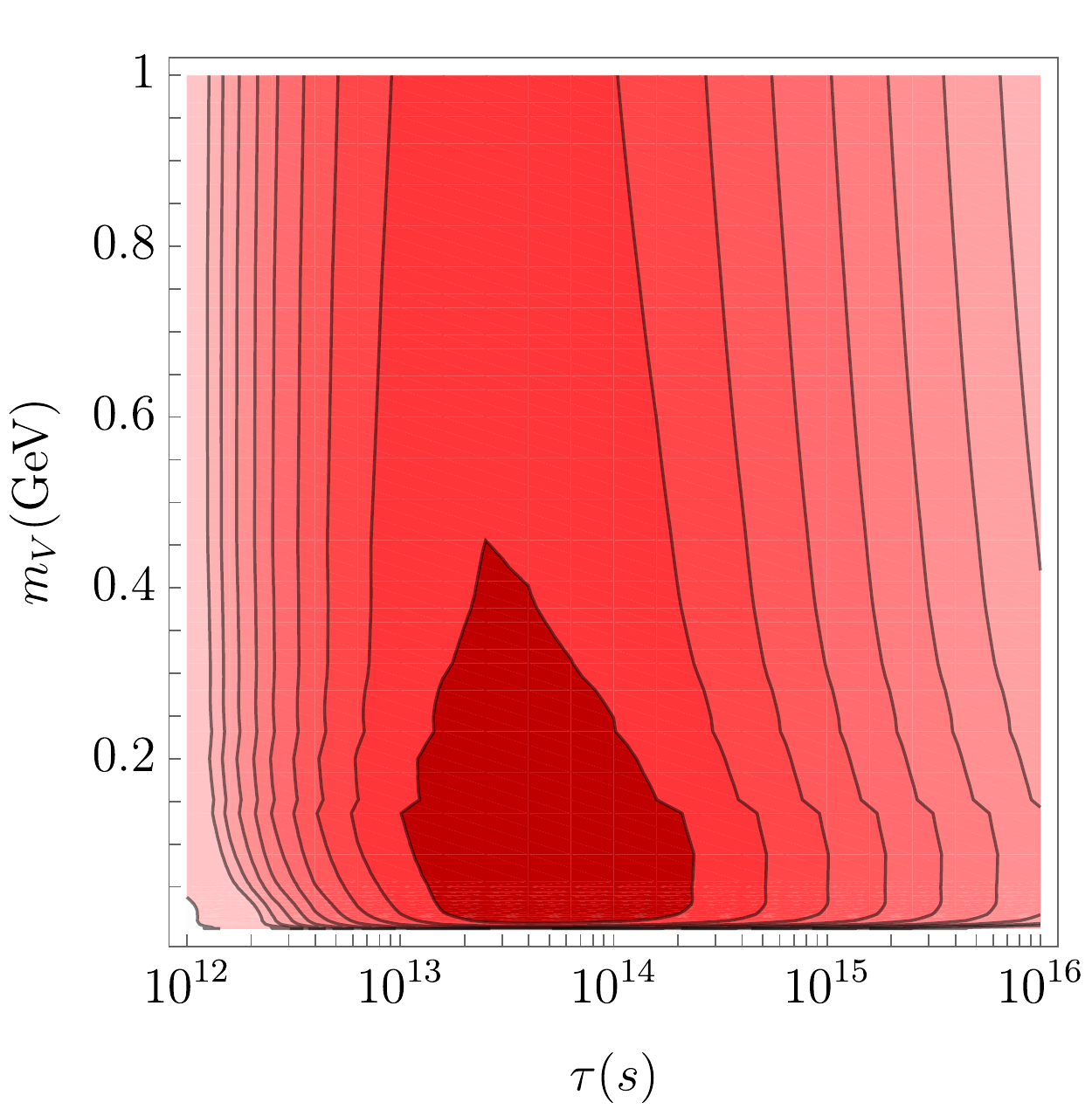}~~
   \includegraphics[width = 0.33\textwidth]{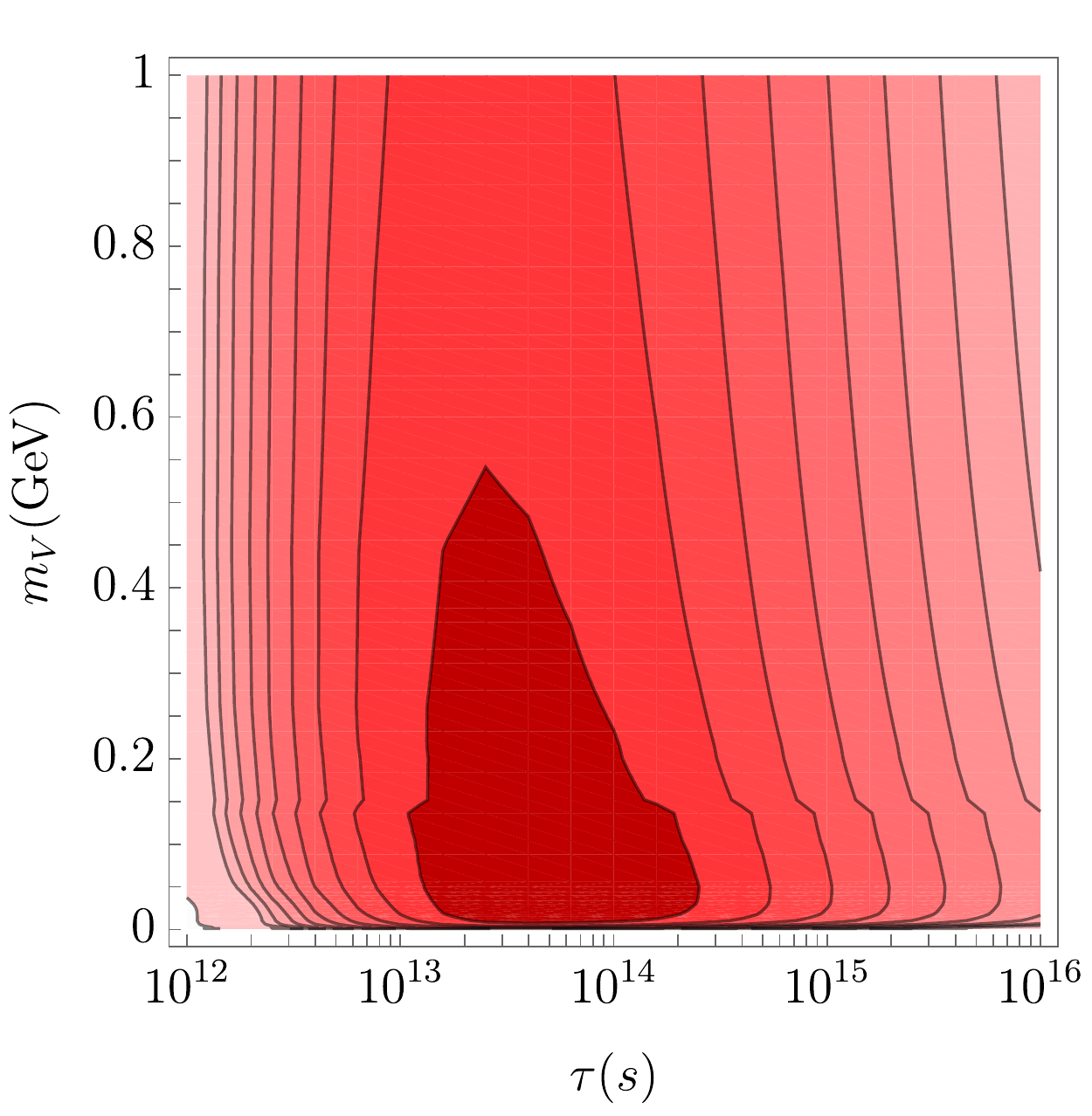}~~
   \includegraphics[width = 0.33\textwidth]{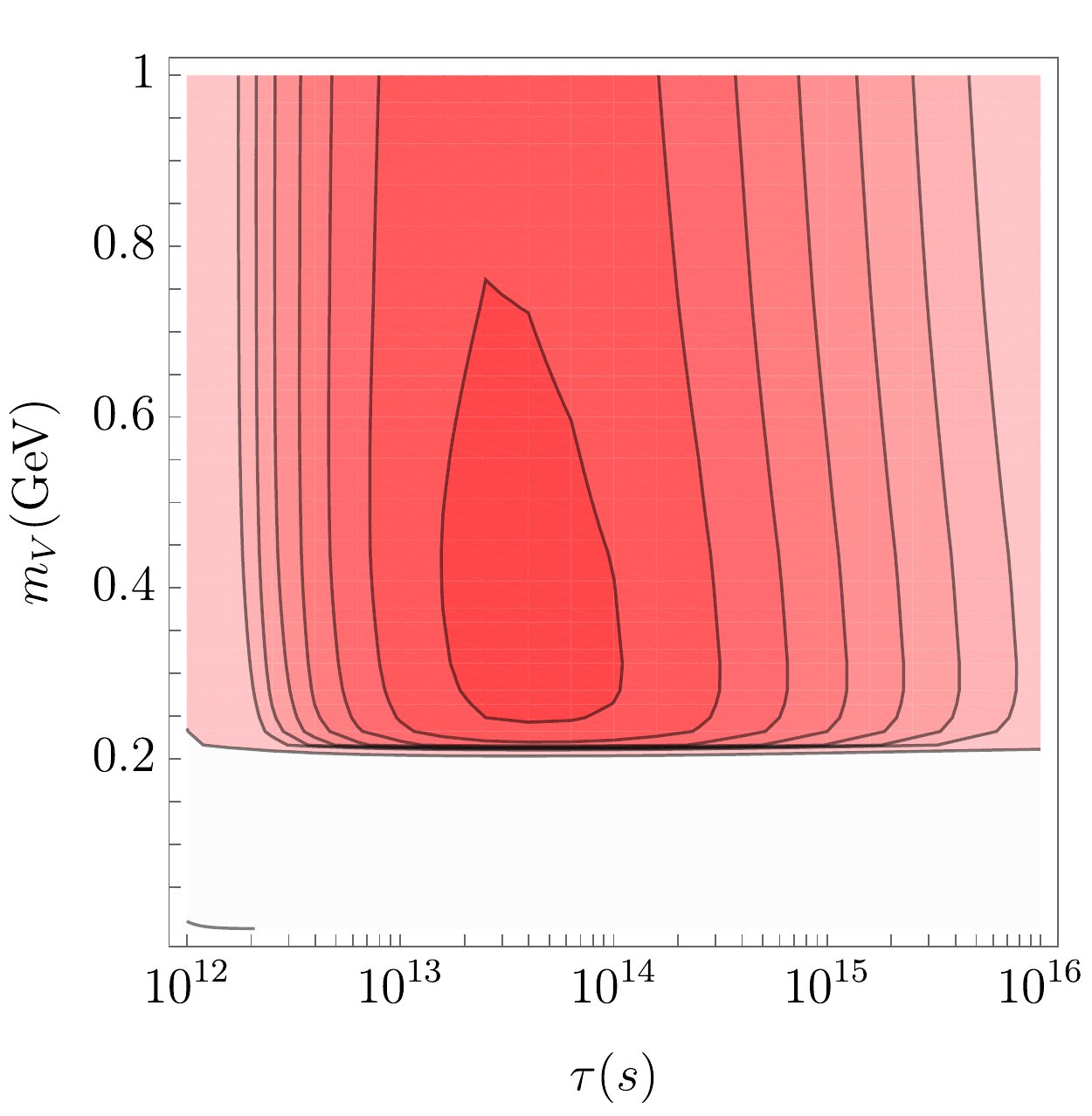}
 \end{center}
\vspace{-0.7cm}
 \caption{
Estimated CMB upper bounds from Planck on the mass times yield $\log_{10}(m_VY_V/\gev)$ 
of sub-GeV dark vectors in the lifetime-mass plane for the dark vector species 
DP~(upper left),
$\vbl$~(upper middle),
$\vem$~(lower left),
$\vet$~(lower middle),
and
$\vmt$~(lower right).
}
 \label{fig:cmblim}
 \end{figure}

  In Fig.~\ref{fig:cmblim} we show the estimated bounds from
deviations in the CMB power spectrum relative to Planck measurements
on the pre-decay dark vector
yield $Y_V = n_V/s$ as a function of decay lifetime $\tau_V$ and mass $m_V$
for the dark vector species
DP~(upper left),
$\vbl$~(upper middle),
$\vem$~(lower left),
$\vet$~(lower middle),
and
$\vmt$~(lower right).
We note that these bounds constrain values of $m_VY_V$ that are
orders of magnitude lower than from BBN.  However, the CMB constraints
fall off quickly for lifetimes below $\tau_V \lesssim 10^{13}\,\text{s}$
and BBN becomes more important.
Let us also emphasize that the results shown for 
$\tau_V \lesssim 10^{13}\,\text{s}$ have a very significant theoretical
uncertainty that we do not attempt to estimate 
(see also Refs.~\cite{Poulin:2016anj,Finkbeiner:2011dx,Slatyer:2016qyl}).
The shape of the CMB bounds mirrors those from BBN, 
with direct photon and electron injection having a greater impact than muons 
or charged pions, and thus the emergence of muon decays and hadronic resonances
with increasing dark vector mass produce distinct features in the figures.

\begin{figure}[ttt]
  \begin{center}
    \includegraphics[width = 0.47\textwidth]{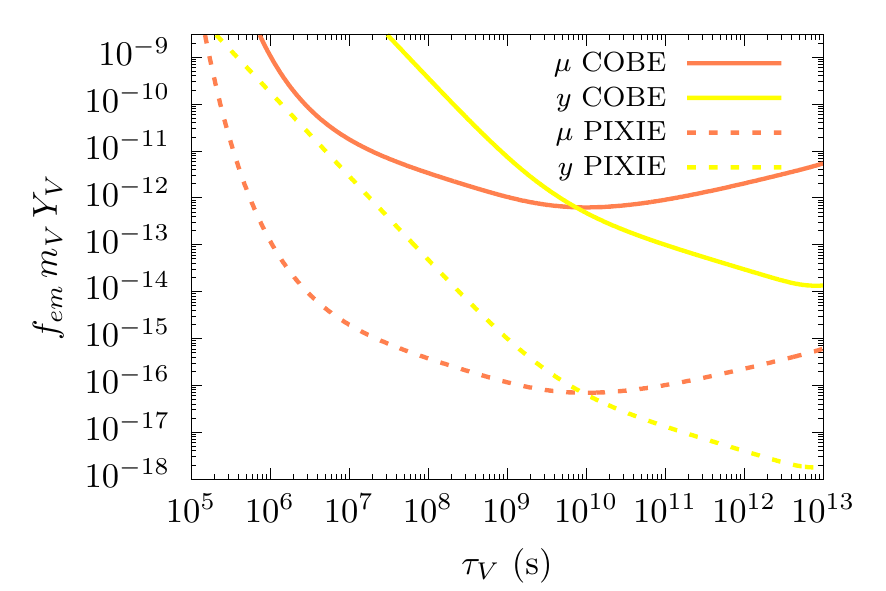}~~
    \includegraphics[width = 0.47\textwidth]{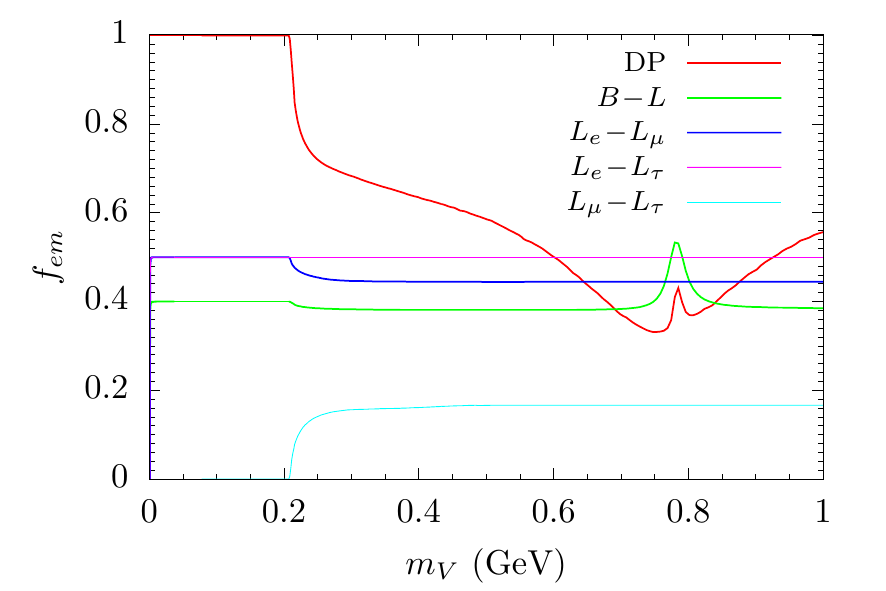}
  \end{center}
  \vspace{-0.7cm}
  \caption{In the left panel we show current (projected) bounds 
from COBE/FIRAS (PIXIE) on the electromagnetic decay factor $f_{em}m_VY_V$ 
from late-time vector decays as a function of the decay lifetime $\tau_V$.  
The right panel shows the electromagnetc decay fraction $f_{em}$ 
for the five dark vector theories studied in this work.}
  \label{fig:cmbf}
\end{figure}

In the left panel of Fig.~\ref{fig:cmbf} we show the approximately 
mass-independent bounds on the combination $f_{em}m_VY_V$ from 
$\mu$- and $y$-type distortions of the CMB frequency spectrum
from COBE/FIRAS measurements as well as projected constraints from PIXIE.
The $\mu$-type distortions are dominant prior to the freezeout of
double-Compton scattering, and $y$-type distortions become more
important after that.  To convert the curves in the left panel
into bounds on a specific dark vector theory, it is only a matter of
determining $f_{em}$.  In the right panel of Fig.~\ref{fig:cmbf} we
plot $f_{em}$ as a function of mass $m_V \in [\mev,\gev]$
for the five dark vector varieties discussed above.
Relative to BBN electromagnetic effects, dark vector decays are constrained
slightly more weakly by current data on the CMB frequency spectrum,
but will become much more constraining with data from PIXIE.  
However, for $\tau_V \gtrsim 10^{13}\,\text{s}$, energy injection primarily alters
the CMB power spectra rather than the frequency 
spectrum~\cite{Zhang:2007zzh,Slatyer:2009yq}, 
and the power spectra give the strongest limits.

\section{Cosmological Limits on Thermal Dark Vectors\label{sec:cosmo}}

  Dark vector bosons can be created through a number of mechanisms in
the early universe, including direct thermal production from SM collisions,
decays or annihilations of other dark sector particles, or in reheating
after inflation. In this section we investigate the thermal production 
of dark vectors since it provides the (nearly) lowest possible dark vector 
population and thus the most model-independent bounds on them.

\subsection{Thermal Production by Freeze-In}

  A minimal population of dark vectors will be created by thermal 
$N\to 1$ freeze-in reactions of the form $(\text{SM})^N \to V$.\footnote{
The only assumption in this statement is that the latest period of radiation
domination once had a temperature $T \gg m_V$ with no significant
entropy injection for $T \lesssim m_V$.}
This production is described by 
\beq
s\frac{dY_V}{dt} = \mathcal{C}[f_V] \ ,
\eeq
where $s$ is the entropy density, $f_V$ is the distribution function of
the dark vector species, 
and $\mathcal{C}[f_V]$ is the standard collision term~\cite{Kolb:1990vq}.  
For underpopulated dark vectors with $f_V \ll f_V^{eq}$ and neglecting
Bose enhancement and Pauli blocking factors, detailed balance implies
\beq
\mathcal{C}[f_V] ~=~ 3\int\!\frac{d^3p_V}{(2\pi)^3}
\lrf{m_V}{E_V}
f_V^{eq}(p_V)\,
\Gamma_V ~=~ \left<\frac{1}{\gamma}\right>n_{V}^{eq}\Gamma_V \ ,
\label{eq:coll1}
\eeq
where $\Gamma_V$ is the total decay width in the $V$ rest frame
and the brackets refer to the thermal average of the inverse Lorentz factor.

The collision term for dark photon production was computed including
thermal corrections and full Fermi-Dirac statistics 
in Ref.~\cite{Fradette:2014sza}.
There it was found that full calculation via leptons (or perturbative quarks)
is approximated very well by evaluating the collision term with 
a Maxwell-Boltzmann distribution and neglecting thermal corrections.
Using a Maxwell-Boltzmann form for $f_V^{eq}$ in Eq.~\eqref{eq:coll1},
the integral can be evaluated analytically with 
the result~\cite{Gondolo:1990dk,Edsjo:1997bg}
\beq
\mathcal{C}[f_V] ~=~ \frac{3}{2\pi^2}\,\Gamma_V\,m_V^2\,T\,K_1(m_V/T) \ ,
\eeq
where $K_1$ is the modified Bessel function of the first kind.

  To evaluate the thermal yield with this collision term, we follow
Ref.~\cite{Fradette:2014sza} and convert time to $x=m_V/T$ and divide 
the production into temperatures above and below the QCD phase transition,
assumed to occur at $\Lambda_{\text{QCD}} \simeq 157\,\mev$.  The net yield is then
\beq
Y_V = (Y_V)_I + (Y_V)_{II} \ ,
\label{eq:yv1}
\eeq
with
\beq
\left(Y_V\right)_I &=& 
\frac{3}{2\pi^2}\,
m_V^3\,\tilde{\Gamma}_V\,
\int_0^{x_{\text{QCD}}}\!dx\;\frac{K_1(x)}{x^2\,s\,H}
\label{eq:yv2}\\
\left(Y_V\right)_{II} &=& 
\frac{3}{2\pi^2}\,
m_V^3\,{\Gamma}_V\,
\int_{x_{\text{QCD}}}^{\infty}\!dx\;\frac{K_1(x)}{x^2\,s\,H}
\ ,
\label{eq:yv3}
\eeq
where $x_{\text{QCD}} = m_V/\Lambda_{\text{QCD}}$ and $\tilde{\Gamma}_V$ 
is the vector decay width into perturbative quark (and lepton) final states.  
For this, we consider up quarks with $m_u = 2.2\,\mev$, 
down quarks with $m_d=4.7\,\mev$, and strange quarks
with $m_s = 93\,\mev$~\cite{Tanabashi:2018oca}.  With this approach,
our results for the yield of the dark photon agree well 
with Ref.~\cite{Fradette:2014sza}.
To the extent that the number of relativistic degrees
of freedom is constant during production and is dominated 
by $T < \Lambda_{\text{QCD}}$,
the final dark vector yield is approximately~\cite{Fradette:2014sza}
\beq
Y_V ~\simeq~ \frac{9}{4\pi}\,\frac{m_V^3\Gamma_V}{(sH)_{x=1}} \ .
\eeq
We use the full result of Eqs.~(\ref{eq:yv1},\ref{eq:yv2},\ref{eq:yv3})
in the analysis to follow.

\subsection{Cosmological Bounds on Thermal Dark Vector Parameters}

By combining the thermal production yields with the BBN and CMB constraints
derived previously, we can put limits on dark vectors produced thermally
in the early universe.  To compare the limits on various dark vector theories 
on an equal footing and to connect with previous work, 
it is convenient to define an effective coupling $\epsilon_{eff}$ by
\beq
\epsilon_{eff} = \left\{
\begin{array}{ccl}
\epsilon&;&\text{dark photon}\\
&&\\
\sqrt{\alpha_V/\alpha}&;&\text{direct-coupling vectors}
\end{array}\right.
\eeq
where $\alpha^{-1} \simeq 137.036$ is the usual fine-structure constant.

\begin{figure}[ttt]
 \begin{center}
   \includegraphics[width = 0.47\textwidth]{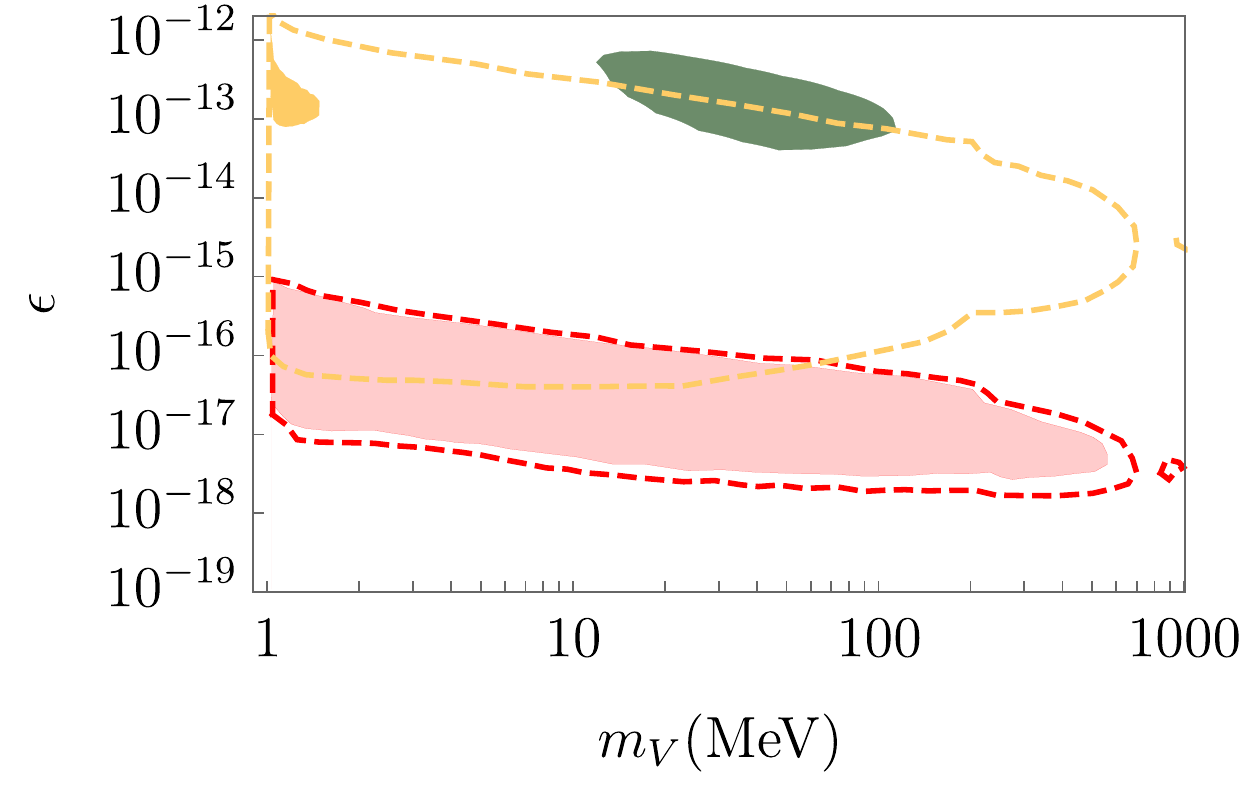}~~
   \includegraphics[width = 0.47\textwidth]{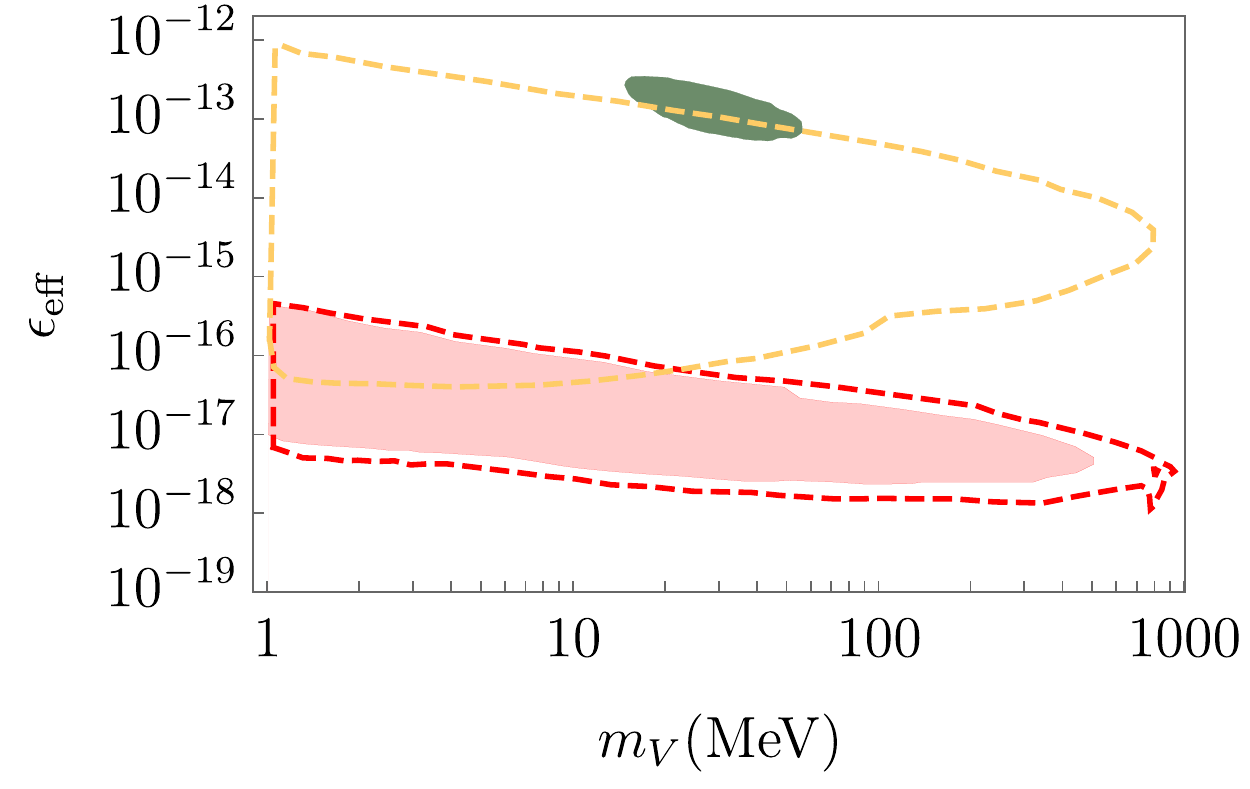}\\
   \includegraphics[width = 0.47\textwidth]{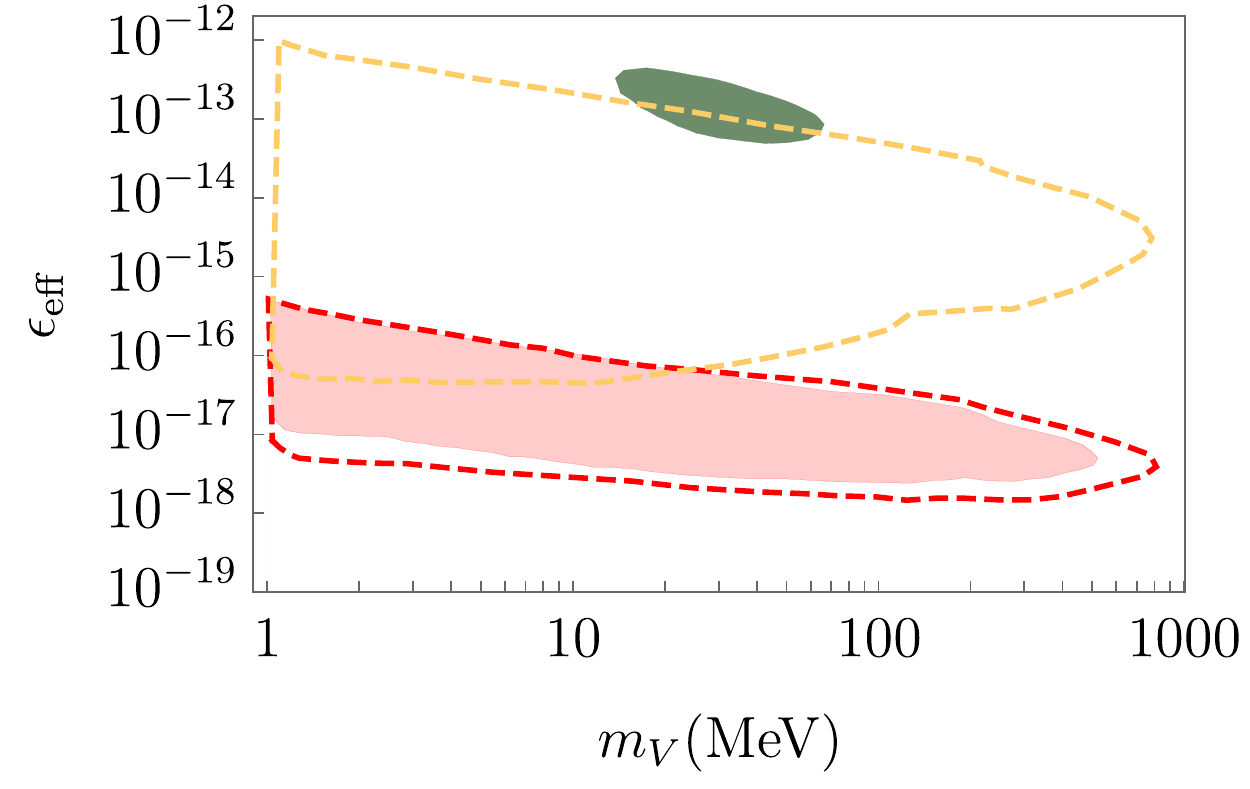}~~
   \includegraphics[width = 0.47\textwidth]{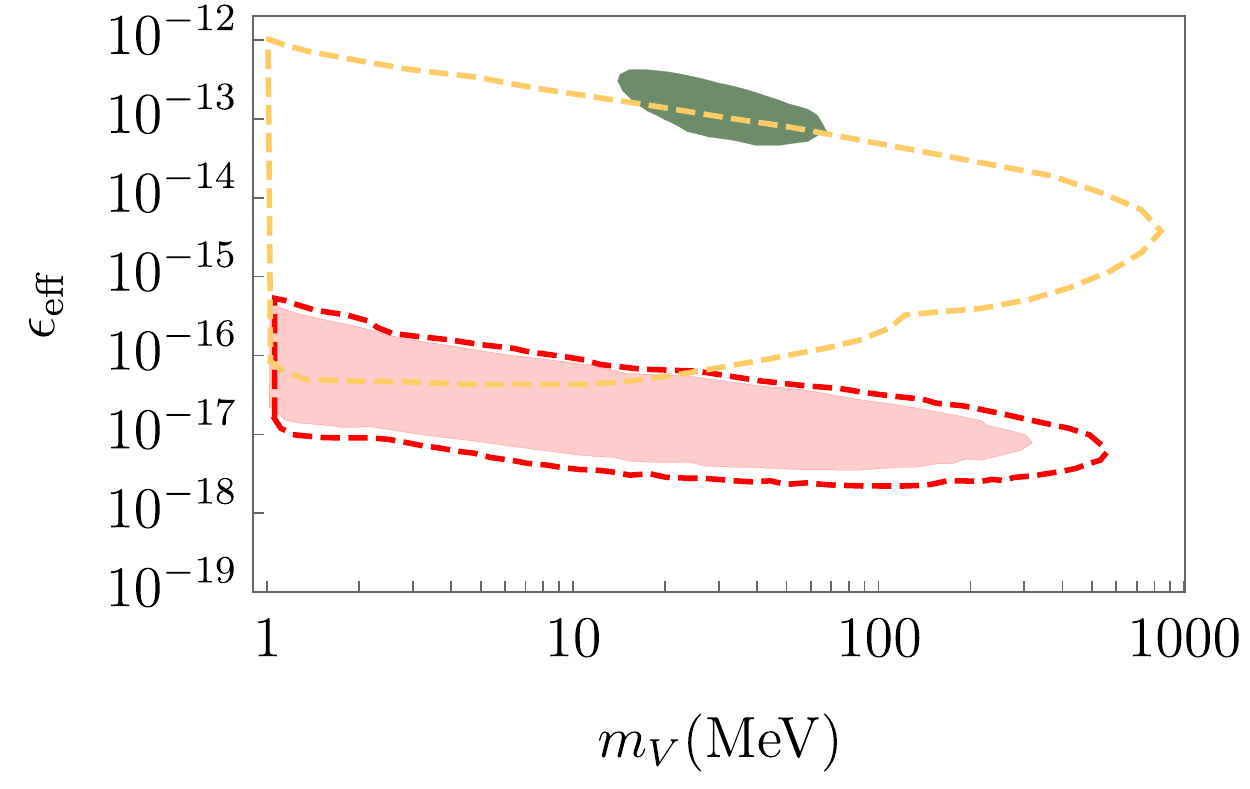}
\end{center}
\vspace{-0.7cm}
 \caption{Cosmological limits on dark vectors from electromagnetic
energy injection as a function of the dark vector mass $m_V$ and 
effective coupling $\epsilon_{eff}$ assuming a thermal freeze-in
abundance.  The dark vector varieties shown are
a dark photon~(upper left),
$\vbl$~(upper right),
$\vem$~(lower left),
and $\vet$~(lower right).
The shaded green regions indicate the exclusions from BBN due to photodissociation, the shaded red regions show the exclusion from the CMB power spectrum measured by Planck, the dashed red contours indicate the projected limits for a cosmic variance limited experiment, the solid yellow region shows the exclusion from CMB frequency spectrum distortions in COBE/FIRAS, and the dashed yellow contours give the projected sensitivity to such distortions at PIXIE.
}
 \label{fig:dvparam}
 \end{figure}

  In Fig.~\ref{fig:dvparam} we show the BBN, CMB-power, and CMB-frequency
exclusions from electromagnetic energy injection in the $m_V$--$\epsilon_{eff}$ 
plane on thermally-produced dark vector species.  
The four panels correspond to the dark vector species
dark photon~(upper left),
$\vbl$~(upper right),
$\vem$~(lower left),
and $\vet$~(lower right), all with dominant decays to the SM.
The green shaded regions show the BBN exclusions from photodissiciation,
the shaded red region gives the bound from Planck deviations in the
CMB power spectra, the dashed red contour indicates the potential future
sensitivity from a cosmic variance limited experiment up to $\ell = 2500$,
the solid yellow region shows the exclusion from COBE/FIRAS from
distortions in the CMB frequency spectrum,
and the dashed yellow contour shows the projected reach of PIXIE for
such distortions.
No significant exclusion is found for the $\vmt$ vector.
Let us also emphasize that these exclusions are conservative in the
sense that they only rely on the assumption of an early universe
dominated by SM radiation at temperature $T\gtrsim m_V$.  The exclusions
would be even stronger if there were other production sources for
the dark vector such as particle decay or annihilation~\cite{Hooper:2012cw,Hasenkamp:2012ii,Hufnagel:2018bjp}.

  The cosmological limits we find for the dark photon are qualitatively
similar to those derived in Refs.~\cite{Fradette:2014sza,Berger:2016vxi}.
Compared to these works, our analysis includes several updates that affect 
the detailed quantitative results for the dark photon and our study extends
to other types of dark vectors.  The most significant update is our treatment 
of electromagnetic energy injection and the resulting
photon cascade spectrum relevant for BBN.  In Ref.~\cite{Fradette:2014sza}
the universal photon spectrum was used, while in Ref.~\cite{Berger:2016vxi}
the modification of the photon spectrum from $e^+e^-$ injection
due to the Thomson limit of inverse Compton scattering was included
leading to much weaker exclusions from BBN photodissociation effects.
Our result lies between these two exclusions -- we confirm the suppression 
of the photon cascade spectrum from lower-energy $e^+e^-$ injection 
pointed out in Ref.~\cite{Berger:2016vxi} but we also find a significant
(and sometimes dominant) additional contribution from final-state photon radiation
in these decays that bring our result closer to Ref.~\cite{Fradette:2014sza}.  
In our treatment of deviations in the CMB power spectrum,
we also use the updated treatment of the ionization fraction presented
in Refs.~\cite{Slatyer:2015kla,Galli:2013dna,Madhavacheril:2013cna,Slatyer:2016qyl}.

  The parameter bounds presented in Fig.~\ref{fig:dvparam} are the strongest
constraints on the species of (thermal) sub-GeV dark vectors studied here
with $\epsilon \lesssim 10^{-12}$ and $m_V \lesssim \gev$. 
For $\epsilon \sim 10^{-11}$ and $m_V \sim \gev$,
Refs.~\cite{Fradette:2014sza,Berger:2016vxi} also found disjoint
BBN exclusions on the dark photon from neutron-proton converstions 
induced by charged pion and kaon injection, 
while for $\epsilon \sim 10^{-13}$ and $m_V \gtrsim 2\,\gev$
they obtained BBN bounds mainly from the direct injection of
neutrons from vector decays.  These BBN bounds from hadronic injection 
are expected to carry over similarly to the $\vbl$ vector for masses above 
$m_V \gtrsim 1\,\gev$. Dark vectors can also alter the effective number
of relativistic neutrino species $N_{eff}$ by thermalizing light singlet 
neutrinos~\cite{Heeck:2014zfa} or by injecting energy preferentially
into either electromagnetic species or neutrinos after neutrino 
decoupling~\cite{Ho:2012ug,Boehm:2013jpa,Nollett:2013pwa,Berlin:2019pbq,Ibe:2019gpv}, with the corresponding bounds based on freeze-in production of a dark
photon extending down to $\epsilon \lesssim 5\times 10^{-10}$ for 
$m_V \lesssim 8\,\mev$.  The emission of gamma-rays by the decays of
dark vectors that would have been produced in supernova 
SN1987A~\cite{Kazanas:2014mca,DeRocco:2019njg}
was shown to rule out couplings down to $\epsilon_{eff} \sim 10^{-12}$ for masses
between 1--100\,MeV~\cite{DeRocco:2019njg}, independent of the cosmological
production mechanism of the dark vector.  Taken together, our cosmological
exclusions of dark vectors (assuming freeze-in production) are disjoint
from these others.

Beyond the limits shown above, we have also estimated the gamma-ray signals 
of dark vector decays with lifetimes longer than the age of the universe.  
Translating these signals to the equivalent from the decay of a long-lived
particle making up all the DM, they correspond to effective DM lifetimes 
greater than $\tau_{eff}\gtrsim 10^{30}\,\text{s}$, well beyond the current limits
of $\tau \gtrsim 10^{25}\,\text{s}$ for $\text{DM}\to e^+e^-$ in this
mass region~\cite{Essig:2013goa}.

\section{Conclusions\label{sec:conc}}

 In this work we have investigated the cosmological bounds on 
a range of sub-GeV dark vectors based on the electromagnetic energy
injected by their decays in the early universe.  This injection
can alter the SM+$\lcdm$ predictions for the light element abundances
from BBN and the power and frequency spectra of the CMB.  Our work
expands on previous studies of cosmological bounds on kinetically-mixed
dark photons~\cite{Fradette:2014sza,Berger:2016vxi}, 
and extends them to four other dark vector species: 
$\vbl$, $\vem$, $\vet$, and $\vmt$.

  The cosmological bounds we derive are based on electromagnetic energy injection.
We take into account deviations from universality in electromagnetic
effects on BBN from lower-energy injection by computing explicitly
the development of the electromagnetic cascade photon spectrum.
In doing so, we take into consideration the detailed photon and electron
energy injection spectra including radiative effects such as FSR.
For the specific case of a dark photon created through thermal freeze-in,
the BBN bounds we find are slightly weaker than Ref.~\cite{Fradette:2014sza} 
and somewhat stronger than Ref.~\cite{Berger:2016vxi}.  Our results for
CMB exclusions on the dark photon appear to be in agreement with 
Refs.~\cite{Fradette:2014sza,Berger:2016vxi}, although we apply a more detailed
energy injection spectrum and we update the effective ionization fraction
following Ref.~\cite{Madhavacheril:2013cna,Slatyer:2015jla,Slatyer:2016qyl}.  
We also extend these methods to other dark vector species.

  Relative to direct laboratory tests of dark vectors, the bounds we find
from cosmology are able to constrain much smaller couplings to SM matter,
with CMB bounds extending all the way to $\epsilon_{eff} \sim 10^{-18}$.
For the dark photon, it is challenging to obtain kinetic mixings 
this small~\cite{Gherghetta:2019coi}, but they can arise when there is
an approximate charge conjugation~($C$) invariance within 
the dark sector~\cite{DiFranzo:2015nli} or collective symmetries~\cite{Koren:2019iuv}.  
For the other direct coupling vectors, the bounds on the gauge couplings we find from cosmology
are strong enough to be potentially relevant for certain superstring 
swampland considerations, corresponding to the set of consistent 
low-energy effective theories that cannot be embedded into a consistent theory 
of quantum gravity~\cite{Vafa:2005ui,ArkaniHamed:2006dz,Ooguri:2006in}.
In particular, $m_V \sim \gev$ and 
$\epsilon_{eff} = \sqrt{\alpha_V/\alpha} \sim 10^{-18}$ approach the
parameter range covered by the weak gravity conjecture proposed 
in Ref.~\cite{ArkaniHamed:2006dz}.

\section*{Acknowledgements}

We thank
Sonia Bacca, Nikita Blinov, Douglas Bryman, Adam Coogan, David Curtin, 
Barry Davids, Thomas Gr\'egroire, Christopher Hearty, David McKeen, Maxim Pospelov, Adam Ritz,
Aleksey Sher, Tracy Slatyer, and Chih-Liang Wu for helpful discussions.  
DEM and GW thank the Aspen Center for Physics, which is supported by National 
Science Foundation grant PHY-1607611, for their hospitality
while this work was being completed.
This work is supported by the Natural Sciences
and Engineering Research Council of Canada~(NSERC), with DEM 
supported in part by Discovery Grants and LF by a CGS~D scholarship.
TRIUMF receives federal funding via a contribution agreement 
with the National Research Council of Canada.



\appendix

\section{{}\hspace{-0.2cm}Appendix: Departures from EM Universality in BBN\label{sec:appb}}

\begin{figure}[ttt]
 \begin{center}
   \includegraphics[width = 0.39\textwidth]{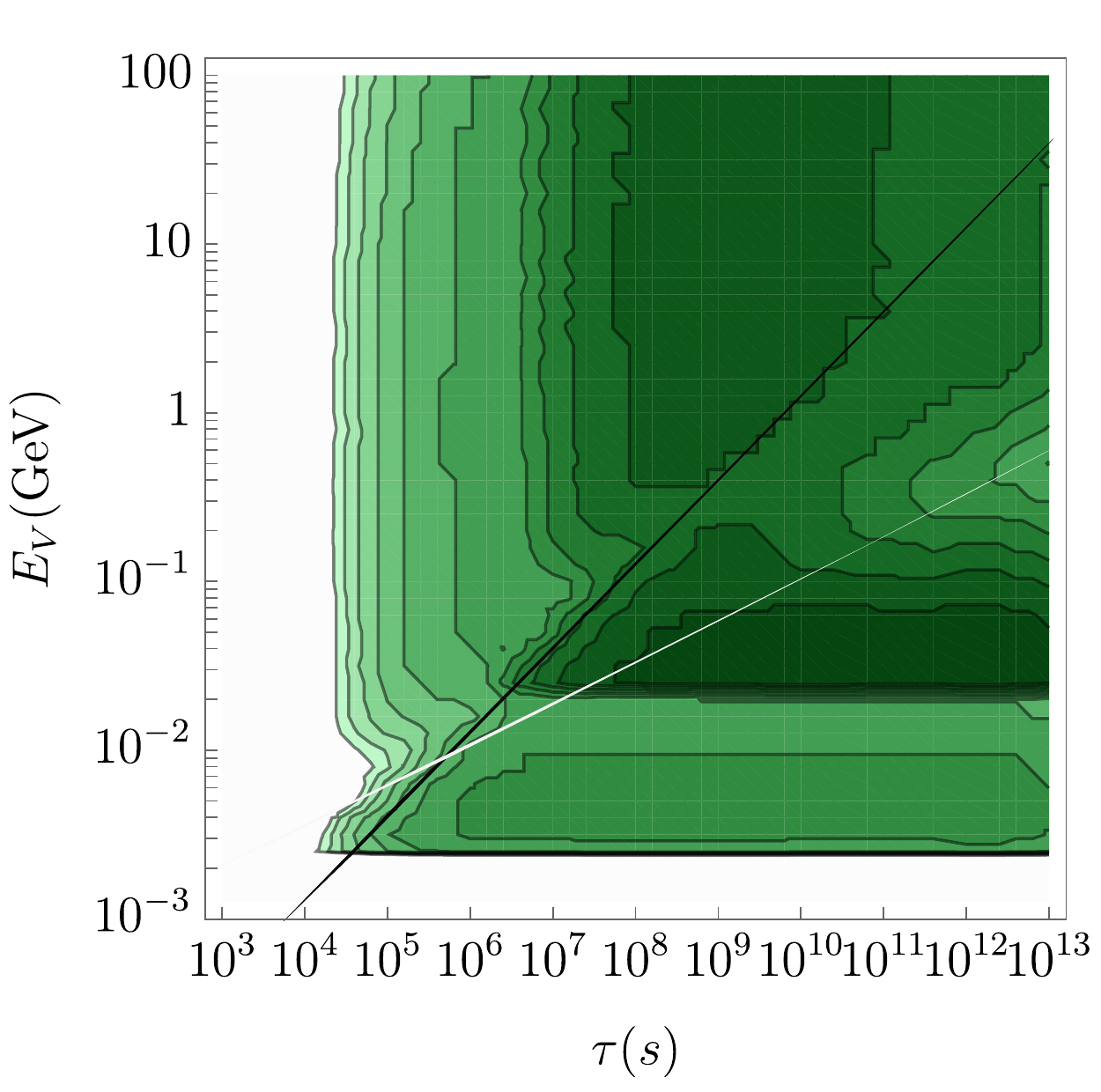}~~
   \includegraphics[width = 0.39\textwidth]{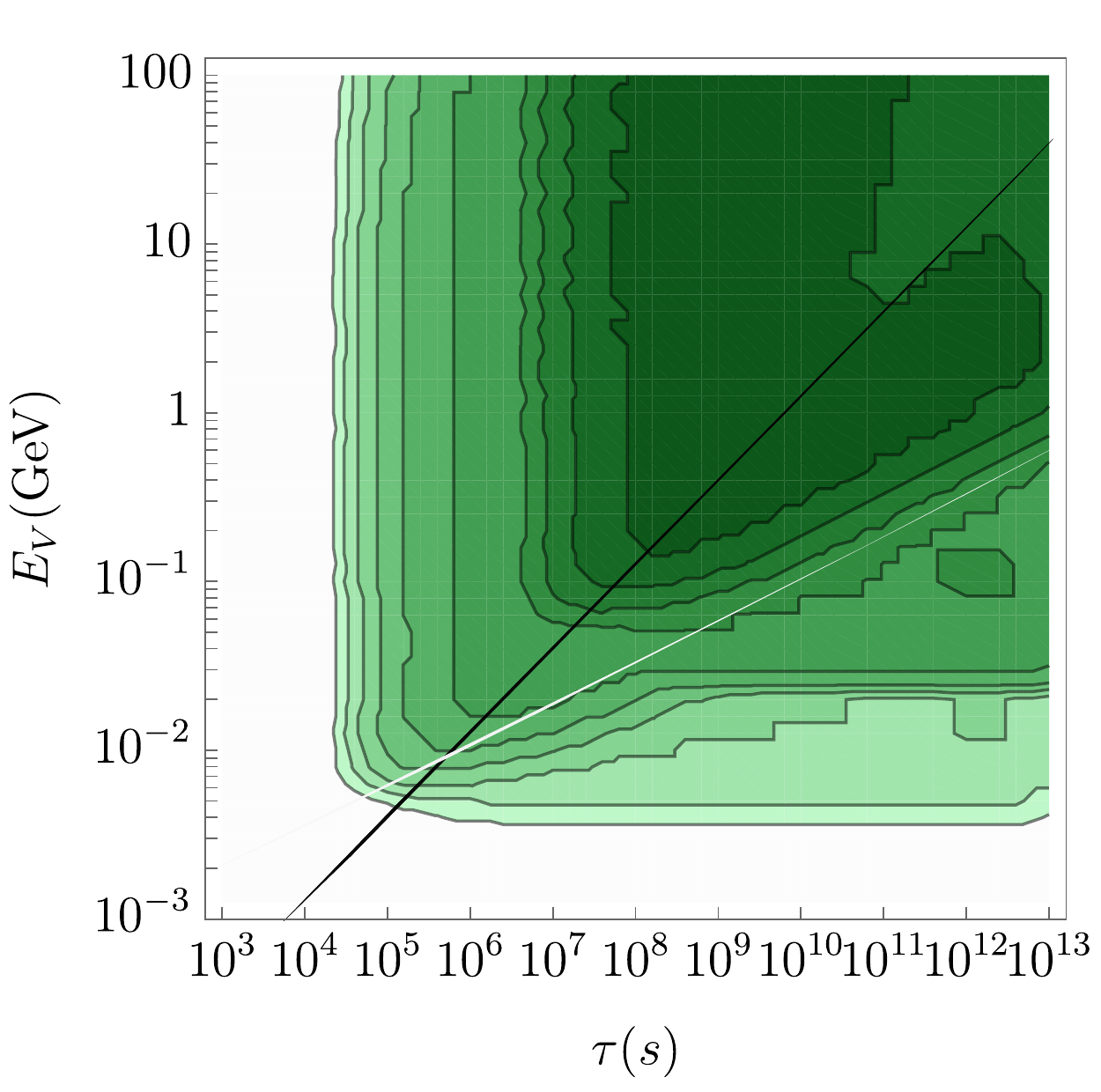}~~
   \includegraphics[width = 0.09\textwidth]{bbn-legend.pdf}
 \end{center}
\vspace{-0.7cm}
 \caption{BBN upper limits on the product $\log_{10}(m_VY_V/\gev)$ as a function of lifetime
$\tau_V$ and injection energy $E_V = m_V/2$ for decays 
$V\to \gamma\gamma$~(left) and $V\to e^+e^-$~(right).  
The black and white lines superimposed on these plots show key
thresholds for the 4P~(black) and IC~(white) processes discussed in the text.}
 \label{fig:bbnuniv}
 \end{figure}

Electromagnetic energy injection well above the weak scale has been shown to
produce a universal photon cascade spectrum that to an excellent approximation
depends only on the total amount of electromagnetic energy 
injected~\cite{Cyburt:2002uv,Kawasaki:1994sc}.
Correspondingly, the limits from BBN on very massive particle decays
depend on the total electromagnetic injection (and decay lifetime) but
not the detailed decay type or energy spectrum.  
In Refs.~\cite{Forestell:2018txr,Poulin:2015woa,Poulin:2015opa,Hufnagel:2018bjp} 
it was shown that this feature breaks down for sub-GeV injections of photons or
electrons, and thus it becomes necessary to track the type of electromagnetic
injection (electrons or photons) and its energy spectrum.

 In Fig.~\ref{fig:bbnuniv} we show the combined bounds from BBN 
on the mass times yield $m_VY_V$ of a massive particle $V$ decaying 
to either $V \to \gamma\gamma$~(left) or $V\to e^+e^-$~(right)
as a function of the lifetime $\tau_V$ and the particle injection
energy $E_V = m_V/2$ from the two-body decay. 
This figure illustrates universality in the
upper left regions of the panels with nearly identical bounds for
photon and electron injection for given values of $\tau_V$ and $m_V$.
However, universality is seen to break down badly in the lower right
regions of the panels, with significantly different BBN bounds depending
on the type of electromagnetic injection within these regions.

The breakdown of universality at lower energies can be understood in terms 
of the energy and temperature dependences of the photon-photon pair 
production~(4P) and inverse Compton~(IC).  These two processes
play a crucial role in the formation of the electromagnetic cascade,
and for high-energy injection they are the dominant processes that seed
the rest of the cascade and drive universality.  As the injection energy 
falls below the weak scale, the typical interaction energy can fall below
important thresholds for these reactions leading to deviations from universality.

Energetic photons injected into the cosmological plasma when the temperature
is below the MeV scale interact primarily with the plasma through the 4P process.
This process is very efficient relative to other processes (and the Hubble rate)
provided the photon energy is greater than a critical energy $E_c$, 
given by~\cite{Cyburt:2002uv,Kawasaki:1994sc}
\beq
E_c = \frac{m_e^2}{22\,T} \ .
\label{eq:ec}
\eeq
This corresponds to the energy threshold for $e^+e^-$ pair production
on background photons extended by a Boltzmann tail.  
For photons with energies above the threshold $E_c$, the 4P
process is extremely efficient and these photons are quickly
reprocessed into electrons at energies below $E_c$. 
In contrast, photons with energy below $E_c$ are only reprocessed
by other, less efficient process.  The black line in Fig.~\ref{fig:bbnuniv}
shows the boundary at which $E_V = E_c(\tau_V)$, with injection energies
above the 4P critical energy to the upper left of this line.

Electrons injected into the cosmological plasma below MeV temperatures
with energy $E_e \gg T$ interact primarily through IC.
This upscatters a background photon and reduces the electron energy.
The energy spectrum of the upscattered photons depends sensitively
on the combination~\cite{Blumenthal:1970gc}
\beq
y_e ~\equiv~ \frac{E_eT}{m_e^2} \ .
\eeq
For $y_e \gg 1$, the scattering occurs in the Klein-Nishina
regime with the outgoing photon energy $E_{\gamma}^{IC}$ typically of the 
same order as the initial electron energy, $E_{\gamma}^{IC} \sim E_e$.
This allows electrons to efficiently populate the photon cascade spectrum.
In contrast, when $y_e \ll 1$ the scattering enters the Thomson regime 
in which the upscattered photon energy is only a small fraction 
of the intial electron energy, and typically less than~\cite{Blumenthal:1970gc}
\beq
E_{\gamma}^{IC} ~\lesssim~ 4\,T\lrf{E_e}{m_e}^2 \ .
\label{eq:iclim}
\eeq
Thus, lower-energy electrons are not able to produce photons through IC 
with enough energy to photodissociate light nuclei.  The white
line in Fig.~\ref{fig:bbnuniv} shows the threshold at which the IC photon
energy produced by an electron with energy $E_e=E_V$ lies below
$E_{\gamma}^{IC} \leq 2\,\mev$, corresponding to the dissociation
threshold for deuterium (and the lowest threshold relevant to our analysis).

  Comparing the BBN exclusion contours in Fig.~\ref{fig:bbnuniv}
to the 4P~(black) and IC~(white) threshold boundaries, we see that 
universality holds to a very good approximation for injection energies
above these lines.  Below these lines, significant deviations from universality
are manifest.  For photon injection with $E_V \sim 20$--$50\,\mev < E_c$,
the photon spectrum is more weakly suppressed and these photons are able
to destroy $^4$He efficiently.  As the photon injection energy falls
below $E_V < 20\,\mev$, the photons are no longer able to dissociate
$^4$He but can still destroy D and $^3$He down to near $2\,\mev$.
These features appear clearly in the left panel.
Electron injection in the Thomson regime, corresponding to the region 
below the white line in the right panel, leads to much weaker limits from 
BBN.  In this region, the dominant contribution to the photon cascade
spectrum is typically hard final-state radiation~(FSR) from the 
primary decay with a relative fraction suppressed 
by a factor of $\alpha$~\cite{Forestell:2018txr}.

\bibliographystyle{JHEP}
\bibliography{ref_bbnvec}

\end{document}